\documentclass[aps,twocolumn,prd,reprint,superscriptaddress,nofootinbib,floatfix]{revtex4-2}
\pdfoutput=1

\usepackage[utf8]{inputenc}
\usepackage{graphicx}
\usepackage{amsmath,amstext, amssymb, amsthm, amsfonts, slashed}
\usepackage{physics}
\usepackage{mathtools}
\usepackage{booktabs}
\usepackage{dcolumn}% Align table columns on decimal point
\usepackage{bm}% bold math
\usepackage[english]{babel}
\usepackage[colorlinks=true,urlcolor=blue,linkcolor=blue,citecolor=blue,hypertexnames]{hyperref}
\usepackage{url}

\usepackage{xspace}
\usepackage{siunitx}
\usepackage{hyperref}
\usepackage[nameinlink]{cleveref}
\usepackage{bookmark}
\usepackage[normalem]{ulem}

\usepackage{xifthen}
\usepackage{xcolor}
\usepackage{enumitem}
\hypersetup{
	colorlinks,
	linkcolor={blue!75!black},
	citecolor={blue!75!black},
	urlcolor={blue!75!black}
}

\graphicspath{{./figures/}}

\begin{document}

%%%%%%%%%%%%%%%%%%%%%%%%%%%%%%%%%%%%
\title{Effective action and black hole solutions in asymptotically safe quantum gravity}

\author{Jan~M.~Pawlowski}
\affiliation{Institut für Theoretische Physik, Universität Heidelberg, Philosophenweg 16, 69120 Heidelberg, Germany}
\affiliation{ExtreMe Matter Institute EMMI, GSI Helmholtzzentrum für Schwerionenforschung mbH, Planckstr.\ 1, 64291 Darmstadt, Germany}

\author{Jan~Tränkle}
%\email[Corresponding author: ]{{jan.traenkle}@{itp.uni-hannover.de}}
\affiliation{Institut für Theoretische Physik, Universität Heidelberg, Philosophenweg 16, 69120 Heidelberg, Germany}
\affiliation{Institut für Theoretische Physik, Leibniz Universität Hannover, Appelstr. 2, 30167 Hannover, Germany}

\begin{abstract}
We derive the quantum effective action and the respective quantum equations of motion from multi-graviton correlation functions in asymptotically safe quantum gravity. The fully momentum-dependent couplings of three- and four-graviton scatterings are computed within the functional renormalisation group approach and the effective action is reconstructed from these vertices. The resulting quantum equations of motion are solved numerically for quantum black hole geometries. Importantly, the black hole solutions show signatures of quantum gravity outside the classical horizon, which manifest in the behaviour of the temporal and radial components of the metric. Three different types of solutions with distinct causal structures are identified and the phase structure of the solution space is investigated.
\end{abstract}

\maketitle

%%%%%%%%%%%%%%%%%%%%%%%%%%%%%%%%%%%%
\section{Introduction}

Observable signatures of quantum gravity are very elusive due to their natural (trans-) Planckian scales. A potential exception is the physics of black holes.
While curvatures outside the horizon of a classical black hole are typically below the Planck scale, the consistent description of these spacetimes in terms of solutions of the quantum equation of motion may still contain traces of quantum gravity in the geometry outside the horizon. 

In the present work we investigate this intriguing scenario within asymptotically safe gravity \cite{Weinberg:1980gg}, where the gravitational interaction approaches an interacting ultraviolet fixed point at trans-Planckian scales.
In the past three decades this quantum field theoretical closure of gravity has been established as a serious contenter for a consistent quantum theory of gravity within the functional renormalisation group (fRG) approach, starting with the seminal paper of Reuter in \cite{Reuter:1996cp}. For recent reviews see \cite{Reuter:2019byg, Eichhorn:2020mte, Bonanno:2020bil, Pawlowski:2020qer}. 

The fRG can be used for computing the momentum-dependent correlation functions of $n$-graviton scatterings, which constitute the expansion coefficients of the effective action of asymptotically safe gravity in an expansion in powers of fluctuating gravitons on a flat background. This fluctuation approach has been set up in \cite{Christiansen:2012rx}, and has been worked out further in \cite{Christiansen:2014raa, Christiansen:2015rva, Denz:2016qks, Bonanno:2021squ} for pure gravity, for reviews see \cite{Reichert:2020mja, Pawlowski:2020qer, Pawlowski:2023gym}. The respective correlation functions allow us to  reconstruct the diffeomorphism-invariant effective action in terms of curvature invariants and covariant momentum dependences. The quantum equations of motion derived from the effective action can be solved for quantum black hole solutions whose geometries outside the horizon are then compared with their classical counterparts in a quest for signatures of quantum gravity.

In \Cref{sec:ReconstructingGamma} we briefly introduce our setup.
In particular, we introduce the quantum effective action for gravity in \Cref{sec:BackgroundEffAction}, and discuss how to reconstruct the contributions of different curvature invariants in terms of momentum-dependent graviton correlation functions in \Cref{sec:DisentanglingInvariants} and \Cref{sec:VertExp}.
In \Cref{sec:fRG} we introduce the functional renormalisation group, and present the computation and results of fully momentum-dependent graviton correlation functions. We use this data to reconstruct the effective action in \Cref{sec:FormFactors}.
In \Cref{sec:QuBHSolutions} we study quantum black hole solutions in our theory. We present the effective field equations, which are the quantum version of Einstein's equations, in \Cref{sec:EffFieldEq}.
Spherically symmetric solutions to the field equations are discussed and classified in \Cref{sec:SphericallySymSol}.
We close with a brief summary in \Cref{sec:Summary}.

%%%%%%%%%%%%%%%%%%%%%%%%%%%%%%%%%%%%
\section{Reconstructing the effective action}
\label{sec:ReconstructingGamma}

In this section we discuss the reconstruction of the diffeomorphism-invariant quantum effective action $\Gamma[g_{\mu\nu}]$.
Because our tool of choice for computing non-perturbative $n$-point correlation functions, the fRG, relies on an ordering of modes in terms of their momenta $p^2>0$ for a successive Wilsonian integrating out of momentum shells, we start by first reconstructing the Euclidean effective action from the correlation functions of the graviton, see e.g.~\cite{Denz:2016qks, Fraaije:2022uhg}. To this end, the effective action is expanded in powers of the curvature up to second order, while retaining the full covariant momentum dependence of the respective terms, that is the dependence on the Euclidean Laplace-Beltrami operator.
Finally, after integrating out all dependencies on the Euclidean cutoff scale, we perform a Wick rotation to obtain the Lorentzian effective action.

%%%%%%%%%%%%%%%%%%%%%%%%%%%%%%%%%%%%
\subsection{The effective action}
\label{sec:BackgroundEffAction}

Euclidean gravity is described by the Wick-rotated Einstein-Hilbert action for the metric field $g_{\mu\nu}$,
\begin{align}
S_{\text{EH}}[g_{\mu\nu}] = \frac{1}{16 \pi G_N}\int_x \sqrt{g} \bigl(R - 2 \Lambda\bigr)\,,
\label{eq:EinsteinHilbertAction}
\end{align}
where $g \coloneqq \det g_{\mu\nu}$, $R$ is the Ricci or curvature scalar, $\Lambda$ denotes the cosmological constant and $G_N=1/M_{\text{pl}}^2$ is the Newton constant related to the Planck mass $M_{\text{pl}}$. \Cref{eq:EffActionwFormFactors} is endowed with a de-Donder-type gauge fixing, see \Cref{sec:GaugeFixing}. 

The quantum counterpart of the classical action \labelcref{eq:EinsteinHilbertAction} is the quantum effective action $\Gamma[\bar g,h]$, and the solution of the equations of motion derived from $\Gamma[\bar g,h]$ governs the full quantum physics. 

In $\Gamma[\bar g,h]$, the full metric $g_{\mu\nu}$ is split into a background metric $\bar{g}_{\mu\nu}$ and dynamical fluctuations $h_{\mu\nu}$, 
\begin{align}
g_{\mu\nu} = \bar{g}_{\mu\nu} + \sqrt{Z_h\, G}\, h_{\mu\nu}\,,
\label{eq:LinearSplit}
\end{align}
where $Z^{1/2}_h$ is the wave function of the transverse-traceless (tt) mode $h_{tt}$ of the fluctuation field $h_{\mu\nu}$. This wave function captures the anomalous scaling of $h_{tt}$ relative to that of the background metric. The dimensional factor $G^{1/2}$ with momentum dimension $[G]=-2$ is related to the Newton constant and leads to the canonical dimension $[h_{\mu\nu}]=1$ of the fluctuation field. In classical gravity we choose $Z_h=1$ and $G=G_N$. Accordingly, the $n^{\text{th}}$-order fluctuation vertices are proportional to $G_N^{(n-2)/2}$.

We also remark that the wave functions of the different graviton modes, the transverse-traceless, the vector and scalar modes, are different from each other, as they receive different quantum corrections.
Hence, it might be advantageous to split the fluctuation term in \labelcref{eq:LinearSplit} into a sum over all modes with roots of their respective wave functions. However, in the present approximation we shall assume the dominance of the transverse-traceless mode and assign its wave function and coupling factor also to the other modes, hence using a uniform $Z_h=Z_{h_{tt}}$. 

Finally, it is convenient to use fully (covariant) momentum-dependent $Z_h$ and $G$. This choice eliminates respective factors in the relations between the background and fluctuation vertices, the Nielsen identities, which is discussed in detail in the following \Cref{sec:VertExp}. Accordingly, we consider 
\begin{align}
Z_h=Z_h(\Delta)\,, \quad  G=G(\Delta)\,,
\end{align}
with
\begin{align}
	\Delta=-{g}^{\mu\nu} {\nabla}_\mu {\nabla}_\nu\,,
	\label{eq:LaplaceOperator}
\end{align}
where we used the covariant derivative $\nabla_\mu$. In the present work, we expand the full effective action in powers of the curvature, that is about flat space. In flat Euclidean space with %
\begin{align}
g_{\mu\nu}=\eta_{\mu\nu} = \delta_{\mu\nu}\,,
\end{align}
the spectral values of the Laplace-Beltrami operator are simply given by the momenta squared, ${ \Delta(\eta) \exp\{\textrm{i} p x\}= p^2  \exp\{\textrm{i} p x\} }$. 

The gauge fixing leads to rather complicated Slavnov-Taylor identities for the correlation functions of the dynamical fluctuation field. However, a diffeomorphism-invariant action is given by the background effective action, defined at vanishing fluctuation field, 
\begin{align}
	\Gamma[g]= \Gamma[g,h=0]\,, 
\label{eq:BackGamma} 
\end{align}
where the background field is identified with the full (mean) metric. It can be shown via the Nielsen identities that the diffeomorphism invariance of the background effective action \labelcref{eq:BackGamma} is the physical one \cite{Pawlowski:2020qer}. 
The Nielsen identities also link background correlation functions to fluctuation field ones, and in the following we reconstruct the background effective action \labelcref{eq:BackGamma} from fluctuation correlation functions.

To that end, we employ a curvature expansion about a flat background and parametrise $\Gamma[{g}]$ with an ansatz of the form
\begin{align}\nonumber 
\Gamma[g_{\mu\nu}]=&\,\frac{1}{16 \pi} \int_x \sqrt{g} \,\Bigg[ \,
{\mathcal{R}}(\Delta , R)\\[1ex]
&\, + R f_{R^2}(\Delta)R +R_{\mu\nu} f_{R_{\mu\nu}^2}(\Delta) R^{\mu\nu} \, \Bigg] \,.
\label{eq:EffActionwFormFactors}
\end{align}
The two two-point correlation functions of the curvature scalar and Ricci tensor, $f_{R^2}(\Delta)$ and $f_{R_{\mu\nu}^2}(\Delta)$, represent generalised dispersions and are also called form factors. These correlation functions or form factors have been investigated in e.g.~\cite{Denz:2016qks} and \cite{Koshelev:2016xqb, Knorr:2018kog, Bosma:2019aiu, Knorr:2019atm, Draper:2020bop, Draper:2020knh, Knorr:2021iwv, Knorr:2021niv, Knorr:2022kqp, Platania:2023uda}, for a recent review see \cite{Knorr:2022dsx}. 

The first term ${\mathcal{R}}(\Delta,R)$ contains the (classical) linear curvature scalar term. This term could be absorbed into the second term with  $f_{R^2}(\Delta)\to f_{R^2}(\Delta, R)$. For the purpose of the present work we prefer to keep it separate, in order to emphasise the underlying expansion in powers of curvature invariants. However, note that it is not possible to simply separate the classical term $\int \sqrt{g} \ G_N^{-1} R$ with a constant Newton coupling $G_N$, as asymptotic safety does not admit such a term in the scale-invariant ultraviolet regime due to the non-vanishing mass dimension of the Newton constant, ${[G_N]=-2}$. The asymptotically safe scaling requires $1/G_N\propto  \Delta$. Promoting the Newton coupling to a function, $G_N\to G(\Delta)$ with $G(0)=G_N$, reduces the respective curvature term to the classical one,  
\begin{align}
	\int_x \sqrt{g}   \frac{1}{ G(\Delta)} R  \simeq \int_x \sqrt{g}   \frac{1}{ G(0)} R\,, 
\label{eq:GN0}
\end{align}
up to potential boundary terms. This is readily seen by expanding $G(\Delta)^{-1} = G(0)^{-1} + c_1 \Delta + ...$, using partial integration and the metric-compatibility of the Levi-Civita connection.
This means that the naive generalisation of $G_N$ to a function of the Laplacian is not suitable for our purpose of constructing a consistent, UV complete effective action.
Accordingly, in the following we will construct an operator ${\mathcal{R}}$ that satisfies infrared (IR) and ultraviolet (UV) consistency, 
\begin{align}\nonumber 
\text{IR:} \quad {\mathcal{R}}(\Delta,R) & \xrightarrow{G_N \Delta \rightarrow 0} \frac{1}{G_N} R \,,\\[1ex]
\text{UV:} \quad {\mathcal{R}}(\Delta,R) & \xrightarrow{G_N\Delta \rightarrow \infty} 0\,,
\label{eq:IRUVlimitsR}
\end{align}
where the asymptotic IR and UV regimes are defined in terms of the spectral values of the Laplace-Beltrami operator $\Delta$, measured in units of the classical Planck mass ${ M_\textrm{pl}^2 = 1/G_N }$.

%%%%%%%%%%%%%%%%%%%%%%%%%%%%%%%%%%%%
\subsection{Reconstructing curvature invariants}
\label{sec:DisentanglingInvariants}

The contributions of different tensor structures can be disentangled by studying the momentum dependence of graviton $n$-point correlation functions $\Gamma^{(n)}(p)$. In \cite{Denz:2016qks} it was found that near the UV fixed point, the RG flow of the three- and four-point functions of the fluctuating graviton $h$ only contains $p^0$, $p^2$ and $p^4$ contributions for $p^2\lesssim k^2$. Here, the powers in $p$ such as $p^{2n}$ should be understood as powers of $(p^2)^n$ and $k$ denotes the cutoff scale of the fRG introduced below. The Ricci tensor $R_{\mu\nu}$ and scalar $R$ both contain two derivatives and thus in momentum space give contributions proportional to $p^2$. Consequently, a $p^4$ term indicates the presence of operators like $R^2$ and $R_{\mu\nu}R^{\mu\nu}$ in the effective action.

For our purposes it is sufficient to only consider tensor structures generated by the local invariants $R^2$ and ${R_{\mu\nu}^2\coloneqq R_{\mu\nu}R^{\mu\nu}}$. The square of the Riemann tensor $R_{\alpha\beta\mu\nu}R^{\alpha\beta\mu\nu}$ can be eliminated by the Gauß-Bonnet term $E[g]$, whose spacetime integral constitutes a topological invariant. Moreover, in $d=4$, the square of the Weyl tensor $C_{\alpha\beta\mu\nu}C^{\alpha\beta\mu\nu}$ can also be expressed in terms of $R^2$, $R_{\mu\nu}^2$ and $E[g]$. In general, also even higher contributions like $p^6$, corresponding to operators like $R^3$, can be present. They were found to be subdominant in previous studies \cite{Denz:2016qks, Falls:2014tra, Falls:2018ylp, Baldazzi:2021orb}, and will therefore not be considered here.

To distinguish the two possible $p^4$-contributions of $R^2$ and $R_{\mu\nu}^2$, we project on the transverse-traceless part of the $n$-point functions. The $R^2$ operator does not contribute to the tt-part of the graviton propagator and the graviton three-point function \cite{ Denz:2016qks, Christiansen:2016sjn, Knorr:2021niv}. Only for the four-point function and higher there is a non-zero overlap. In turn, the overlap of $R_{\mu\nu}^2$ tensor structures with the tt-projected three- and four-point functions is non-vanishing. This implies that any $p^4$-contribution to $\Gamma^{(3)}$ can be unambiguously attributed to an $R_{\mu\nu}^2$ term. Moreover, the $p^4$ term in $\Gamma^{(4)}$ can then be exploited to extract the contribution of $R^2$ tensor structures.

In summary, this structure allows us to reconstruct the form factors $f_{R^2}(\Delta)$ and $f_{R_{\mu\nu}^2}(\Delta)$  of the $R^2$ and $R_{\mu\nu}^2$ terms in the effective action \labelcref{eq:EffActionwFormFactors} from the momentum dependence of the three- and four-point function.

%%%%%%%%%%%%%%%%%%%%%%%%%%%%%%%%%%%%%%%
\subsection{Vertex expansion in the fluctuation approach} 
\label{sec:VertExp}

\begin{figure}[t]
	\centering
	\includegraphics[trim= 80 10 70 10, width=0.9\columnwidth]{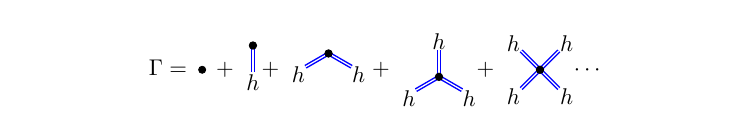}
	\caption{Schematic representation of the vertex expansion of the effective action $\Gamma[\bar{g},h]$. Blue double lines represent the fluctuating graviton and black dots mark full vertices.}
	\label{fig:VertexExpansion}
\end{figure}

In order to compute the form factors, we employ the vertex expansion in the fluctuation approach to gravity \cite{Christiansen:2012rx, Christiansen:2014raa, Christiansen:2015rva, Denz:2016qks, Bonanno:2021squ}, for reviews see \cite{Reichert:2020mja, Pawlowski:2020qer}. In this expansion scheme we expand the full quantum effective action $\Gamma[\bar{g},h]$ in powers of the fluctuation field $h$.
Schematically, this expansion can be written in diagrammatic form as displayed in \Cref{fig:VertexExpansion}.
The technical aspects of the expansion are detailed in \Cref{app:VertexExpansion}. 

In the present work we utilise the fact that the non-trivial terms in the Nielsen identities drop in significance in the higher orders, for a respective discussion see \cite{Pawlowski:2020qer}. With the linear split \labelcref{eq:LinearSplit}, this leads us to 
\begin{align}
\Gamma^{(n)}[g]=\Gamma^{(n,0)}[g,0]\approx  \left( Z_h G\right)^{-n/2}\Gamma^{(0,n)}[g,0]\,, 
\label{eq:NIbargh}
\end{align}
for the one-particle irreducible (1PI) correlation functions 
\begin{align}
\Gamma^{(n,m)}[\bar g,h]=  \frac{\delta^{n+m}\Gamma[\bar g, h ]}{\delta \bar g^n\,\delta h^m}\,.
\end{align}
\Cref{eq:NIbargh} relates the expansion of the effective action $\Gamma[\bar g, h ]$ in terms of the fluctuation field $h_{\mu\nu}$ to that of the diffeomorphism-invariant action $\Gamma[g]$ in terms of the full metric $g_{\mu\nu}$.

%
%%%%%%%%%%%%%%%%%%%%%%%%%%%%%%%%%%%%
The content of \labelcref{eq:NIbargh} is best illustrated at the example of the transverse-traceless component of the graviton two-point function in a flat background with $\Delta \to p^2$ with the parametrisation  
\begin{align} \nonumber 
	\Gamma_{\textrm{tt}}^{(2,0)}(p)  =&\,\frac{1}{16 \pi G_N} \,Z_g(p)\, p^2\,\Pi_{\textrm{tt}} \,,\\[1ex]
	\Gamma_{\textrm{tt}}^{(0,2)}(p)  =&\,\frac{1}{16 \pi} Z_h(p) \,p^2\,\Pi_{\textrm{tt}} \,. 
\label{eq:Gamma2}
\end{align}
The tt-projection operator $\Pi_\textrm{tt}$ is provided in \labelcref{eq:TT_projector} in \Cref{app:ProjectionScheme}. We have normalised the wave function of the background graviton with ${Z_g(p=0)=1}$, which fixes the Newton constant as the infrared or classical one. Inserting \labelcref{eq:Gamma2} into \labelcref{eq:NIbargh} leads us to 
\begin{align} 
	G(p) = \frac{G_N}{Z_g(p)}\, ,
\label{eq:GNZg}
\end{align}
which fixes the prefactor $G(p)$ in \labelcref{eq:LinearSplit} as the inverse wave function of the transverse-traceless background graviton, $Z_g$, multiplied by the classical Newton constant $G_N$.

\Cref{eq:GNZg} encodes the well-known relation that the inverse wave function of the background field in the background field approach carries the running coupling of the theory. Indeed, \labelcref{eq:GNZg} should rather be seen as the definition of a running Newton coupling for the reasons explained around \labelcref{eq:GN0}. 

Note that in terms of the diffeomorphism-covariant correlation functions $\Gamma^{(n)}$, the wave function $Z_g(p) $ receives contributions from different diffeomorphism-invariant terms in the effective action. In the approximation used here, it is a weighted sum of contributions or (two-point) couplings of the $\cal R$- and $R_{\mu\nu}^2$-terms in \labelcref{eq:EffActionwFormFactors}. Accordingly, while the inversion of $Z_g(p)$ makes sense for defining an average two-point function coupling, for higher correlation functions we first have to disentangle the different invariants in the sum before doing an inversion of the parts.  

While this can be done with a sum over general tensor structures, we now restrict ourselves to that used in the present work: On the left hand side of \labelcref{eq:NIbargh} we use results for the completely tt part of the vertices at the momentum symmetric point, and parametrise the $n$-point functions as
\begin{align}
	\Gamma_{\textrm{tt}}^{(n)}(\boldsymbol{p}) &=\gamma^{(n)}_{g}(p) \, \mathcal{T}_{R,\textrm{tt}}^{(n)}(\boldsymbol{p})\,.
	\label{eq:gammanBack}
\end{align}
where $\boldsymbol{p}=(p_1,...,p_n)$, $p= \vert p_i \vert$ and $\mathcal{T}_{R,\textrm{tt}}^{(n)}$ is the completely transverse-traceless projection of the $n^{\text{th}}$-derivatives of the curvature part of the Einstein-Hilbert action, see  \labelcref{eq:TTPartSEHn} in \Cref{app:ProjectionScheme}. There it is also detailed how to project $\Gamma^{(n)}$ on the tt-coefficient $\gamma_g^{(n)}$, see \labelcref{eq:gammanBackApp}. Note, that due to our choice of the tensor structures the vertex coefficients $\gamma_g^{(n)}$ have momentum dimension two, ${[\gamma_g^{(n)}]=2}$, in analogy with the classical theory where $[G_N^{\phantom{N}-1}]=2$.

In the present work we compute the fluctuation vertices $\Gamma^{(0,n)}$ and use the approximated Nielsen identity \labelcref{eq:NIbargh} to extract the coefficients $\gamma_g^{(n)}$ of the $n$-point functions of the diffeomorphism-invariant action \labelcref{eq:BackGamma} from these results. Accordingly, we can already parametrise the $\Gamma^{(0,n)}$ in terms of $\gamma_g^{(n)}$, to wit, 
\begin{align}
	\Gamma_{\textrm{tt}}^{(0,n)}(\boldsymbol{p}) &=\, \prod_{i=1}^{n}\sqrt{Z_h(p_i) \,G(p_i)  }\, \gamma_g^{(n)}(p) \,\mathcal{T}_{R,\textrm{tt}}^{(n)}(\boldsymbol{p})\,.
	\label{eq:gammanFluc}
\end{align}
The projection of the full $\Gamma^{(0,n)}$ onto $\Gamma_{\textrm{tt}}^{(0,n)}$ is detailed in  \Cref{app:ProjectionScheme}, see \labelcref{eq:gammanFlucApp}. 

We note in this context, that the vertex couplings, which genuinely drive the dynamics in the fluctuation approach, are not the $\gamma_g^{(n)}$, but rather the coupling coefficients $\gamma_h^{(n)}(\boldsymbol{p})$ given by  
\begin{align}
	\gamma_h^{(n)}(\boldsymbol{p})=  \prod_{i=1}^{n}\sqrt{G(p_i) } \gamma_g^{(n)}(\boldsymbol{p}) \,. 
\label{eq:Gn}
\end{align}
These are the coefficients of the RG-invariant core $\bar\Gamma^{(0,n)}$ of the fluctuation $n$-point functions, 
\begin{align} 
	\bar\Gamma^{(0,n)}(\boldsymbol{p})  = \frac{\Gamma^{(0,n)}(\boldsymbol{p}) }{ \prod_{i=1}^{n}\sqrt{Z_h(p_i)  }}\,.
\label{eq:barGn}
\end{align}
\Cref{eq:barGn} naturally occurs in diagrams as all fluctuation propagators carry a $1/Z_h$, and a factor $1/Z_h^{1/2} $ can be assigned to each vertex these propagators are connected to.
In short, the loop diagrams in functional approaches, which are built from full vertices and propagators, reduce to diagrams with classical propagators and $\bar\Gamma^{(0,n)}$ vertices.
Hence, these vertices carry the full quantum dynamics of the theory.  Finally, \labelcref{eq:barGn} is simply the non-perturbative analogue of the standard perturbative definition of the running coupling from vertices and propagators.
By now, the relevance as well as significance of \labelcref{eq:barGn} in non-perturbative expansion schemes has been evaluated and shown in both, QCD and asymptotically safe gravity, for respective reviews see \cite{Dupuis:2020fhh, Pawlowski:2020qer, Fu:2022gou}. 

In order to clarify the meaning of the yet undetermined coefficient $G(p)$, as defined in \labelcref{eq:GNZg}, we use that both, $\Gamma_\textrm{tt}^{(2)}$ and $\Gamma_\textrm{tt}^{(3)}$, only receive contributions from $\cal R$ and the Ricci tensor part of the effective action \labelcref{eq:EffActionwFormFactors}.
Hence, in a last approximation step we identify $\gamma_g^{(2)}$ and $\gamma_g^{(3)}$, leading to 
\begin{align}
	G(p) = \left[\gamma_h^{(3)}(p)\right]^2= \left[\gamma_g^{(3)}(p)\right]^{-1} \,,
\label{eq:DefofG}
\end{align}
where we have used $\gamma_h^{(2)}=1$. Note also, that the relation \labelcref{eq:DefofG} and similar ones for higher $n$-point functions suggest the definition ${\gamma_h^{(n)} = G_n^{n/2-1} }$. Here, the $G_n$ are referred to as avatars of the Newton coupling, see \cite{Denz:2016qks, Eichhorn:2018akn, Eichhorn:2018ydy, Pawlowski:2020qer}. However, for the present purpose of reconstructing the effective action in terms of curvature invariants this parametrisation is not suggestive and we will not use it here.   

In summary, we are led to a systematic expansion of the full diffeomorphism-invariant action $\Gamma[g]$.
Note that this is a non-perturbative expansion, as all the vertices appearing in 
\Cref{fig:VertexExpansion} are fully dressed. The present expansion \labelcref{eq:EffActionwFormFactors} in powers of the curvature suggests to expand the effective action in terms of correlation functions on a flat Euclidean background
\begin{align}
\bar{g}_{\mu\nu} = \delta_{\mu\nu}\,,
\label{eq:flatbackground}
\end{align}
where all curvature invariants vanish, i.e.~$R=0$ and ${R_{\mu\nu}^2 =0}$.
This simplifies computations considerably, as they can be done in momentum space. Expansions about different backgrounds have been considered e.g.~in \cite{Christiansen:2017bsy, Burger:2019upn}. Promising signs of apparent convergence of the vertex expansion about a flat background have been observed in \cite{Denz:2016qks} upon gradually improving the truncation, by computing the full running vertices up to the four-point function. 

The analysis in the present work is based on a fRG computation of momentum-dependent fluctuation ${\text{two-,}}$ three- and four-graviton correlation functions, $\Gamma^{(0,n)}$ with $n=2,3,4$, building on the computation in \cite{Denz:2016qks}.
While in the latter work the full momentum dependence was already provided implicitly in terms of the ultraviolet-infrared flows, we will integrate these results in the present work.
The fRG computation is described in the following \Cref{sec:fRG}, further details can be found in \cite{Denz:2016qks, Pawlowski:2020qer}.

%%%%%%%%%%%%%%%%%%%%%%%%%%%%%%%%%%%%
\section{Functional renormalisation group flows for gravity}
\label{sec:fRG}

\begin{figure}[t]
  \centering
  \includegraphics[width=0.7\columnwidth]{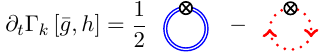}
  \caption{Diagrammatic representation of the Wetterich equation. The blue double line represents the fluctuating graviton, the dotted line denotes the ghost field. The crossed circle indicates the insertion of a regulator derivative, $\otimes  = \partial_t \mathcal{R}_k$.}
  \label{fig:WetterichEqDiag}
\end{figure}

For the computation of the momentum-dependent ${\text{two-,}}$ three- and four-point function we employ the functional renormalisation group, for recent reviews including quantum gravity see \cite{Dupuis:2020fhh, Saueressig:2023irs}.
In this approach, quantum fluctuations are integrated out successively by introducing a momentum-dependent mass term with the mass function $\mathcal{R}_k$ to the classical dispersion. Such a deformation suppresses fluctuations with momenta $p^2 \lesssim  k^2$ below the IR cutoff scale $k$, and leaves the UV-regime with $p^2 \gtrsim  k^2$ unchanged. The flow of the scale-dependent effective action is given by the Wetterich equation \cite{Wetterich:1992yh} for quantum gravity \cite{Reuter:1996cp}, 
\begin{subequations}
\label{eq:FunFlow}
\begin{align}
\partial_t \Gamma_k[\bar{g},h]  =\frac{1}{2}
     \text{Tr}\, \mathcal{G}_k[\bar{g},h] \, \partial_t \mathcal{R}_k\, .
     \label{eq:WetterichEq}
\end{align}
In \labelcref{eq:WetterichEq} we used the definitions ${\partial_t= k\partial_k}$ and $\mathcal{G}$  is the full, field-dependent propagator of the fluctuation fields $\phi=(h_{\mu\nu},c_\mu,\bar c_\mu)$ defined in \labelcref{eq:SuperField}, 
\begin{align}
	\mathcal{G}_k[\bar g,h]  = \frac{1}{\Gamma_k^{(0,2)}[\bar{g},h]+\mathcal{R}_k}\,,
\label{eq:Fullprop}
\end{align}
\end{subequations}
where we drop the ghost field argument, as all our correlation functions are evaluated at vanishing ghost fields, $(c_\mu,\bar c_\mu)=0$. The trace operator Tr incorporates a sum over all fluctuation fields $\phi$, Lorentz indices and an integral over momenta or spacetime, and includes a minus sign for the  Grassmann-valued ghost fields. A graphical depiction is provided in \Cref{fig:WetterichEqDiag}.
%
%%%%%%%%%%%%%%%%%%%%%%%%%%%%%%%%%%%%
%%%%%%%%%%%%%%%%%%%%%%%%%%%%%%%%%%%%
%

%%%%%%%%%%%%%%%%%%%%%%%%%%%
\subsection{Flows for $n$-point functions} 
\label{eq:FlowsnPoinit}

\begin{figure}[!t]
  \centering
  \includegraphics[width=\columnwidth]{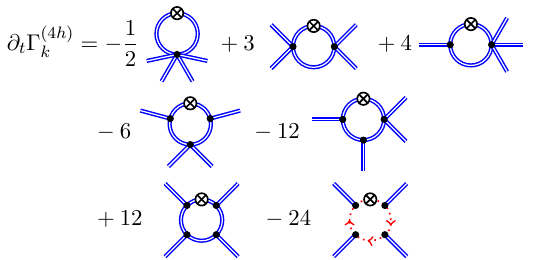}
  \caption{Diagrammatic representation of the flow of the graviton four-point function. All vertices and propagators are fully dressed. Reproduced from \cite{Denz:2016qks}.}
  \label{fig:Flow4PtFct}
\end{figure}

The flows of the 1PI $n$-point correlation functions are derived from the functional flow equation \labelcref{eq:FunFlow} by taking $n$-field derivatives w.r.t.~to the fields $\bar g \, , \phi$. Inserting the parametrisation \labelcref{eq:gammanFluc} of $\Gamma_{\textrm{tt}}^{(0,n)}$ into the flow equation allows one to extract momentum-dependent, non-perturbative beta functions for the couplings $\gamma_h^{(n)}$ defined in \labelcref{eq:Gn}. These flows give rise to UV fixed points and asymptotically safe RG trajectories. Here we build upon the results in \cite{Christiansen:2012rx, Christiansen:2014raa, Christiansen:2015rva}, and in particular \cite{Denz:2016qks, Bonanno:2021squ} where the momentum dependence of fluctuation and background correlation functions has been resolved up to the graviton four-point function. Results for the momentum-dependent propagators and running Newton coupling have been provided directly in \cite{Bonanno:2021squ}, but the momentum dependence of the vertices has only be given implicitly in terms of the flows. Using this existing data, in the present work we integrated the fully momentum-dependent flow of the four-point function, which is depicted diagrammatically in \Cref{fig:Flow4PtFct}. This provides, for the first time, the full momentum dependence of the graviton four-point function $\Gamma^{(4)}(p)$ at the physical cutoff scale $k=0$.
%
%%%%%%%%%%%%%%%%%%%%%%%%%%%%%%%%%%%%
%%%%%%%%%%%%%%%%%%%%%%%%%%%%%%%%%%%%
%

\begin{figure*}[ht!]
  \centering
  \includegraphics[width=0.49\textwidth]{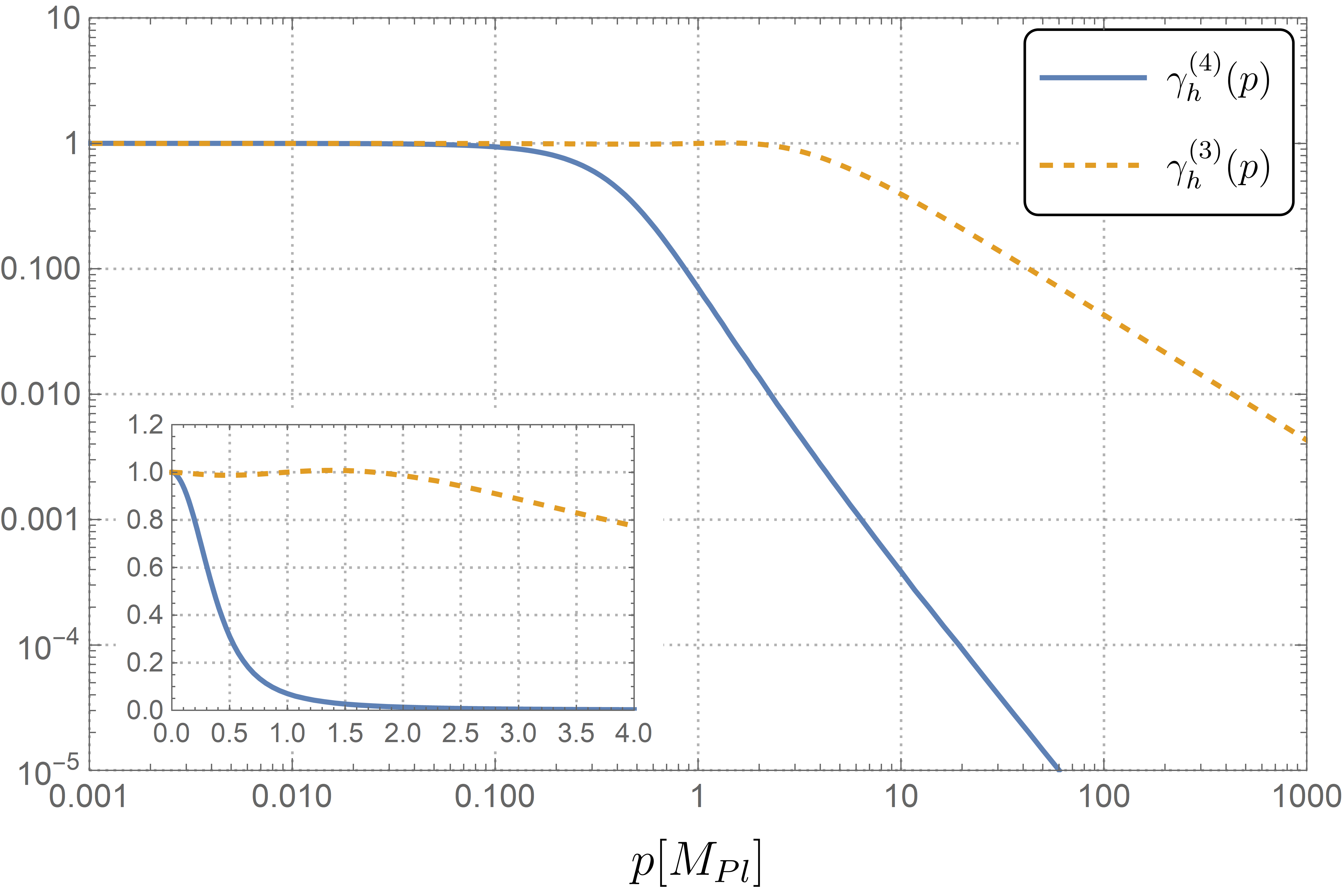}
  \includegraphics[width=0.49\textwidth]{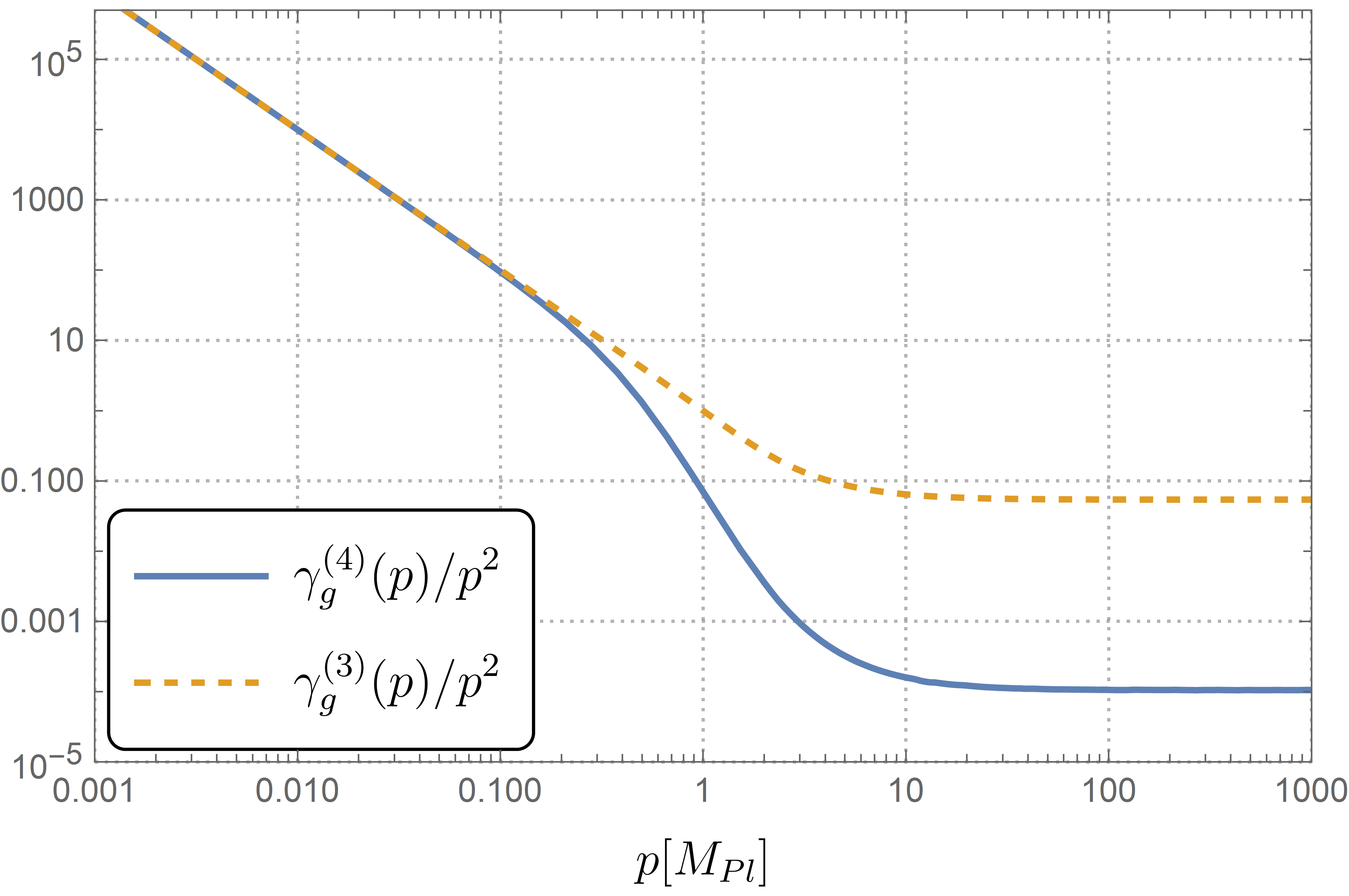}
  \caption{Momentum dependence of the fluctuation (left) and background (right) couplings of the three- and four-graviton vertex in the physical limit $k\rightarrow 0$.
The dimensionful fluctuation couplings $\gamma_h^{(4)}(p)$ (blue) and $\gamma_h^{(3)}(p)$ (dashed) approach a constant value in the IR, given by the Newton coupling $(\gamma_h^{(3)})^2(0)=\gamma_h^{(4)}(0)=G_N\equiv 1$. $\gamma_h^{(4)}(p)$ turns to the asymptotically safe running $\propto p^{-2}$ roughly one order of magnitude earlier than $\gamma_h^{(3)}(p)$, which runs as $\propto p^{-1}$ due to its mass dimension $[\gamma_h^{(3)}]=-1$. The inset shows the same data plotted linearly. The dimensionless background couplings $\gamma_g^{(4)}(p)/p^2$ (blue) and $\gamma_g^{(3)}/p^2$ (dashed) scale $\propto p^{-2}$ in the IR and go to a constant in the UV fixed point regime. The data used for $\gamma_{g,h}^{(3)}(p)$ was computed in \cite{Bonanno:2021squ}.
  }
  \label{fig:MomDepG4}
\end{figure*}

We defer the discussion of our approximations and technical details of the full computation to \Cref{sec:FlowofFourPtFct} and proceed with the discussion of our results, which are summarised in \Cref{fig:MomDepG4}.
%
%%%%%%%%%%%%%%%%%%%%%%%%%%%%%%%%%%%%
%%%%%%%%%%%%%%%%%%%%%%%%%%%%%%%%%%%%
%

There, we display the momentum-dependent fluctuation coupling $\gamma_h^{(4)}(p)$ of the four-graviton vertex at vanishing cutoff $k=0$. It constitutes a key result of this work. We find a classical regime $p\ll  M_{\text{Pl}}$, where the coupling is constant and $\gamma_h^{(4)}(0) = G_N$. Towards the fixed point regime above the Planck scale we see asymptotically safe scaling $\propto p^{-2}$, reflecting the mass dimension ${[ \gamma_h^{(4)} ] = -2}$.
Interestingly, the coupling $\gamma_h^{(4)}(p)$ shows the non-trivial scaling behaviour already roughly one order of magnitude below the Planck scale and before its lower counterpart $\gamma_h^{(3)}(p)$, which is constructed from data computed in \cite{Bonanno:2021squ}. This is due to the fact that the fixed point values of the dimensionless couplings \labelcref{eq:dimlesscouplings} satisfy $g_3^*>g_4^*$. Our finding implies that the quantum regime of asymptotically safe gravity already leads to non-trivial signatures one order of magnitude below the Planck scale. Analytic fits to the numerical data are provided in \Cref{sec:Fits}.

On the right-hand side of \Cref{fig:MomDepG4} we display the dimensionless counterparts $\gamma_g^{(4)}(p)/p^2$ and $\gamma_g^{(3)}(p)/p^2$ of the background couplings, which are computed from the fluctuation couplings by means of the relation \labelcref{eq:Gn}. They show canonical scaling $\propto p^{-2}$ in the classical regime and take constant values in the UV, as required by asymptotic safety.

%%%%%%%%%%%%%%%%%%%%%%%%%%%%%%%%%%%%
\section{Form factors}
\label{sec:FormFactors}

We now discuss the reconstruction of the form factor factors $f_{R^2}(p^2)$ and $f_{R_{\mu\nu}^2}(p^2)$ introduced in \labelcref{eq:EffActionwFormFactors} from the momentum-dependent background couplings $\gamma_g^{(3)}(p)$ and $\gamma_g^{(4)}(p)$ at hand.

In a first step we project the $n$-point functions on the curvature part of the Einstein-Hilbert action by contracting them with $\mathcal{T}_{R,\textrm{tt}}^{(n)}$. This projection of the background $n$-point functions \labelcref{eq:gammanBack} results in the scalar expression 
\begin{align}
\Gamma_{\textrm{tt}}^{(n)} \circ \mathcal{T}_{R,\textrm{tt}}^{(n)}/p^2 = C_{R}^{(n)} \gamma_g^{(n)} p^2 \, ,
\label{eq:dimless-vertex}
\end{align}
where the division with $p^2$ eliminates the dispersion part of the tensor structure and $\circ$ denotes the pairwise contraction of Lorentz indices. The projection scheme is described in more detail in \Cref{app:ProjectionScheme} and the values of the coefficients $C_{R}^{(n)}$ are given in \labelcref{eq:projection_coefficients}.

The projection leading to \labelcref{eq:dimless-vertex} is now also applied to the $n$-point functions resulting from respective field derivatives of \labelcref{eq:EffActionwFormFactors}. Equating the two sides results in
\begin{align}
C_{R}^{(n)} \gamma_g^{(n)} p^2 = C_{\mathcal{R}}^{(n)}g_{\mathcal{R}} p^2 + C_{R^2}^{(n)} f_{R^2} p^4 + C_{R_{\mu\nu}^2}^{(n)} f_{R_{\mu\nu}^2} p^4\,,
\label{eq:DerivingFormFactors}
\end{align}
where the coefficients $C_{\mathcal{O}}^{(n)}$ are given explicitly in \labelcref{eq:projection_coefficients}. These preparations enable us to read-off the form factors. We start with the Ricci tensor form factor, which is obtained for $n=3$,
\begin{align}
f_{R_{\mu\nu}^2}(p^2)= \frac{1}{C_{R_{\mu\nu}^2}^{(3)}} \frac{C_{R}^{(3)} \gamma_g^{(3)}-C_{\mathcal{R}}^{(3)}g_{\mathcal{R}}}{p^2} \, ,
\label{eq:fRmunu_R}
\end{align}
where we used that $C_{R^2}^{(3)}=0$ due to the tt-projection. The $R^2$ form factor is computed from the four-point function, $n=4$. \Cref{eq:DerivingFormFactors} results in 
\begin{align}
f_{R^2}(p^2)= \frac{1}{C_{R^2}^{(4)}} \frac{C_{R}^{(4)} \gamma_g^{(4)}-C_{\mathcal{R}}^{(4)}g_{\mathcal{R}}- C_{R_{\mu\nu}^2}^{(4)}f_{R_{\mu\nu}^2} p^2}{p^2} \,.
\label{eq:fR2_R}
\end{align}
It is left to reconstruct the ${\mathcal{R}}(\Delta,R)$-term, that encodes the classical curvature term in the IR. In order to satisfy the required limits \labelcref{eq:IRUVlimitsR} we choose 
\begin{align}
\mathcal{R}(\Delta,R) = R \frac{\gamma_g^{(3)}(\Delta) -\bar{\gamma}_3 \Delta}{\Delta+ R} R\,,
\label{eq:R_operator}
\end{align}
with the dimensionless constant $\bar{\gamma}_3$ defined in \labelcref{eq:gammabar3} in \Cref{sec:GenEHTerm}. \Cref{eq:R_operator} generates a $p^2$ contribution to the flow of the $n$-point functions ${\Gamma}_G^{(n)}(p)$ with the prefactor given by ${g_{\mathcal{R}} =\gamma_g^{(3)}(0) = G_N^{-1}}$, in accordance with the classical limit. This is shown in \Cref{sec:GenEHTerm}.
The denominator is a sum of the curvature scalar $R$ and the Laplace-Beltrami operator $\Delta$. The latter part arranges for the necessary disappearance of the curvature term in the UV with large spectral values of $\Delta$. In turn, in the IR with small spectral values of $\Delta$, \labelcref{eq:R_operator} reduces to $\gamma_g^{(3)}(0)\,R$. We emphasise that the structure of \labelcref{eq:R_operator} is fixed, but the general denominator in \labelcref{eq:R_operator} is given by a weighted sum of $R$ and $\Delta$. 

\begin{figure}
  \centering
  \includegraphics[width=\columnwidth]{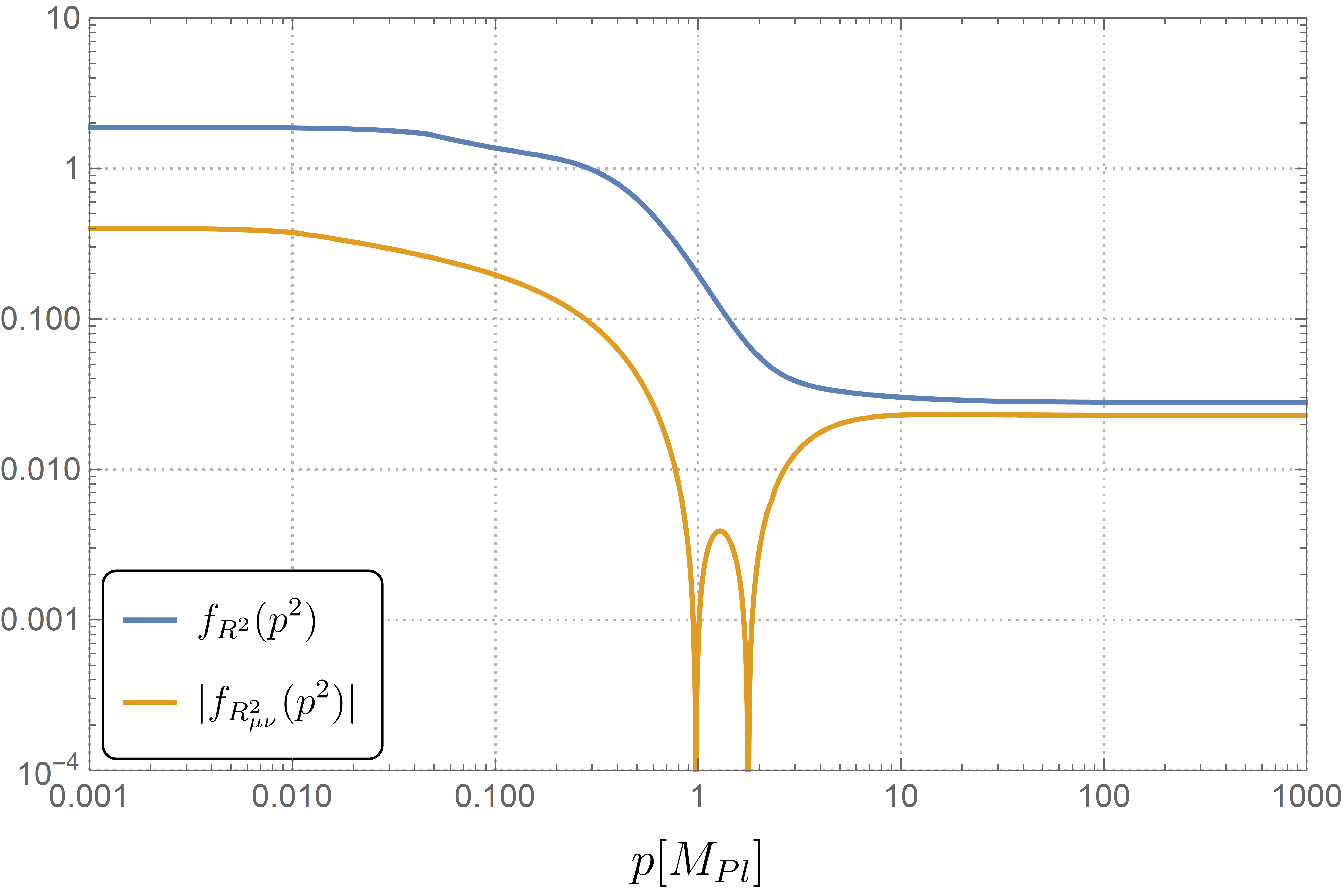}
  \caption{The form factors $f_{R_{\mu\nu}^2}(p^2)$ \labelcref{eq:fRmunu_R} and $f_{R^2}(p^2)$ \labelcref{eq:fR2_R} as a function of physical momentum $p$. Both attain constant values in the IR as well as in the UV. Note that $f_{R_{\mu\nu}^2}$ is negative over almost the entire momentum range, except near $p\approx 1$, and $\vert f_{R_{\mu\nu}^2} \vert < f_{R^2}$ everywhere.}
  \label{fig:FormFactors}
\end{figure}

In \Cref{fig:FormFactors} the form factors $f_{R_{\mu\nu}^2}(p^2)$ and $f_{R^2}(p^2)$ are plotted as a function of the physical momentum $p$.
Both form factors approach constant values in the IR as well as in the UV. The values resulting from our numerical data are given in \Cref{sec:NumericalValues}, \labelcref{eq:IRcouplingsvalues} and \labelcref{eq:UVcouplingsvalues} respectively.
We observe that the UV values are one to two orders of magnitude smaller than their IR counterparts.
Furthermore, we note that $f_{R_{\mu\nu}^2}(p^2)$ is negative both in the IR and UV and its absolute value is smaller than $f_{R^2}(p^2)$ everywhere.

%%%%%%%%%%%%%%%%%%%%%%%%%%%%%%%%%
%%%%%%%%%%%%%%%%%%%%%%%%%%%%%%%%%

For practical applications we derive very simple analytic expressions, that interpolate between the IR and UV asymptotes of  $f_{R_{\mu\nu}^2}$ and  $f_{R^2}$. Substituting ${p^2 \to \Delta}$, in position space these are given by
\begin{align}
 f_{R_{\mu\nu}^2}(\Delta) & \simeq \tilde{g}_{R_{\mu\nu}^2} + \frac{g_{R_{\mu\nu}^2} - \tilde{g}_{R_{\mu\nu}^2}}{1+ p_0^{-2}\Delta} \,,\nonumber \\[1ex]
 f_{R^2}(\Delta) & \simeq \tilde{g}_{R^2} + \frac{g_{R^2} - \tilde{g}_{R^2}}{1+ p_1^{-2}\Delta}\,,
\label{eq:formfactorSimpleAnalytic}
\end{align}
with $p_0$ and $p_1$ expressed in units of the Planck mass $p_i/M_{\text{Pl}}$ and the numerical values of the constants given in \labelcref{eq:IRcouplingsvalues}, \labelcref{eq:UVcouplingsvalues} and \labelcref{eq:fitparametersSimpleAnalytic} respectively.
These expressions capture the correct limits of the form factors as ${\Delta \to 0}$ and ${\Delta \to \infty}$.
More elaborate fits to the numerical data are provided in \Cref{sec:Fits}, see \labelcref{eq:FitFuncfRmn} and \labelcref{eq:FitFuncfR2}.
Importantly, we find that the form factors introduce non-local contributions in the effective action due to the appearance of the inverse Laplace operator $\Delta^{-1}$.

Finally, we study the infrared behaviour of the form factor expressions
\labelcref{eq:fRmunu_R,eq:fR2_R,eq:R_operator,eq:formfactorSimpleAnalytic}.
By construction, the operator $\mathcal{R}(\Delta,R)$ in \labelcref{eq:R_operator} reduces to the Einstein-Hilbert term, see \labelcref{eq:IRUVlimitsR}.
The low energy limit of the form factors is found by Taylor expanding the analytic expressions \labelcref{eq:FitFuncfRmn} and \labelcref{eq:FitFuncfR2} about $p^2=0$ to first order.
In position space this yields
\begin{align}\nonumber
 f_{R_{\mu\nu}^2}(\Delta) & \simeq g_{R_{\mu\nu}^2} + c_1 \Delta \,, \\[1ex]
 f_{R^2}(\Delta) & \simeq g_{{R}^2} + c_2 \Delta\,,
 \label{eq:formfactorTaylor}
\end{align}
with the values of the coefficients $c_{1,2}$ given in \labelcref{eq:TaylorCoefficients}. In terms of the vertex dressings $\gamma_g^{(3)}$ and $\gamma_g^{(4)}$, the constant IR couplings of the $R_{\mu\nu}^2$ and $R^2$ terms read
\begin{align}\nonumber 
g_{R_{\mu\nu}^2} &=  c_3 \left[\gamma_g^{(3)}\right]'(0)\,, \\[1ex]
g_{R^2} &= c_4 \left[\gamma_g^{(4)}\right]'(0)-  c_5 \left[\gamma_g^{(3)}\right]'(0)\,,
\label{eq:gRmunu+gR2}
\end{align}
with the prime denoting a derivative with respect to $p^2$ and the constants $c_{3,4,5}$ defined in \labelcref{eq:IRcoefficients}.

The reconstruction of the effective action, or rather the form factors, is an important result of the present work. It puts forward a systematic way to map the momentum dependent graviton correlation functions, which can be computed non-perturbatively within the fRG, to diffeomorphism invariant expressions in the full quantum effective action. The approximations used for the explicit computation can be lifted within improved computations.
Apart from the uncertainties arising from these technical approximations, our reconstructed (Euclidean) effective action for asymptotically safe gravity is essentially unique and thus, the resulting effective field equations can be used to address important physics questions, as for example in the context of black hole spacetimes.

%%%%%%%%%%%%%%%%%%%%%%%%%%%%%%%%%%%%
\section{Quantum black hole solutions}
\label{sec:QuBHSolutions}

Asymptotic safety dictates a specific form of scale-dependence of the gravitational coupling in the UV, which is expected to ultimately resolve the curvature singularities in the classical theory.
The implications of "quantum-improving" classical black hole spacetimes by promoting the classical Newton constant $G_N$ to a scale-dependent quantity have received a lot of attention \cite{Bonanno:2000ep, Reuter:2006rg, Bonanno:2006eu, Reuter:2010xb, Falls:2010he, Cai:2010zh, Becker:2012js, Falls:2012nd, Koch:2013owa, Koch:2013rwa, Koch:2014cqa, Saueressig:2015xua, Pawlowski:2018swz, Adeifeoba:2018ydh, Held:2019xde, Platania:2019kyx, Borissova:2022jqj, Chen:2022xjk, Borissova:2022mgd, Bonanno:2023rzk}, see \cite{Eichhorn:2022bgu, Platania:2023srt} for recent reviews.
This procedure is expected to capture the leading order quantum effects and interesting implications have been reported.
However, the identification of the IR cutoff $k$ with a physical scale is not unique in gravity and different choices can result in physically distinct spacetimes.
Moreover, this RG improvement can introduce unphysical coordinate dependencies \cite{Held:2021vwd} and physical conclusions have to be drawn with care.
In our novel approach we eliminate the need for this ambiguous scale-identification by first integrating out all fluctuations to obtain $\Gamma[g]$ at $k=0$. Finding solutions to the resulting effective field equations then allows for a fully consistent description of quantum black hole spacetimes.
\\
\\
To capture the causal structure of a given spacetime geometry, it is essential to consider the theory in Lorentzian signature. Therefore, at this point we perform a Wick rotation on our covariant Euclidean results and denote this by replacing the Euclidean Laplace operator by the full, Lorentzian d'Alembertian, $\Delta \to \Box$, in the effective action \labelcref{eq:EffActionwFormFactors} and the expressions for the form factors derived above, $f(\Delta)\to f(\Box)$.
\footnote{On a flat background, the Wick rotation would amount to ${\Delta=-\partial_t^2-\partial_i\partial^i \longrightarrow \Box=+\partial_t^2-\partial_i\partial^i}$.}

Although the Wick rotation in curved spacetime is still not fully understood \cite{Candelas:1977tt, Visser:2017atf, Baldazzi:2018mtl}, there is a growing body of studies which indicate that asymptotic safety is realized also in a Lorentzian setting. This has been shown for example employing the Arnowitt-Deser-Misner (ADM) decomposition of the metric \cite{Manrique:2011jc, Rechenberger:2012dt, Biemans:2016rvp, Knorr:2018fdu, Saueressig:2023tfy}, using spectral reconstructions \cite{Bonanno:2021squ, Fehre:2021eob}, Lorentzian fRG equations \cite{DAngelo:2022vsh, DAngelo:2023wje} and is investigated also in lattice simulations \cite{Ambjorn:2012jv, Ambjorn:2013tki, Ambjorn:2022naa} and with causal sets \cite{Eichhorn:2019xav, deBrito:2023axj}.
The reported phase diagrams and fixed point structure, as well as the UV-scaling behaviour, show striking similarities between the two signatures. Therefore, we may expect that the main physical features, like the existence of the fixed point and the asymptotically safe scaling of the fluctuation couplings, which we rely on for the reconstruction of the form factors, carry over to the physical signature.

%%%%%%%%%%%%%%%%%%%%%%%%%%%%%%%%
\subsection{Effective field equations}
\label{sec:EffFieldEq}

In order to address the question how black holes are modified in the quantum theory, we solve the field equations resulting from the effective action $\Gamma[\bar{g}]$ \labelcref{eq:EffActionwFormFactors}.
In a slight abuse of notation we will write $g_{\mu\nu}$ when we actually refer to the background metric $\bar{g}_{\mu\nu}$ in this section.

If we neglect the $g_{\mu\nu}$ dependence of the form factors themselves, we can straightforwardly derive the contribution to the field equations resulting from the second two terms in \labelcref{eq:EffActionwFormFactors}.
Dropping boundary terms that arise from partial integration, the term ${R f_{R^2}(\Box)R}$ results in the field equations
\begin{align}
2\left( R_{\mu\nu} - \frac{1}{4}Rg_{\mu\nu} + g_{\mu\nu}\Box - \nabla_\mu\nabla_\nu\right) f_{R^2}(\Box)R = 0\,,
\label{eq:RfREoM}
\end{align}
and variation of the term ${R_{\mu\nu} f_{R_{\mu\nu}^2}(\Box) R^{\mu\nu}}$ results in
\begin{align} \nonumber
&\left(2 R_{\mu\sigma}g_{\nu\rho} - \frac{1}{2}g_{\mu\nu}R_{\rho\sigma} +g_{\rho\mu}g_{\sigma\nu}\Box \right)f_{R_{\mu\nu}^2}(\Box)R^{\rho\sigma}  \\[1ex]
&\hspace{.6cm}+\left( g_{\mu\nu}\nabla_\rho\nabla_\sigma  - 2 g_{\mu\sigma}\nabla_\rho \nabla_\nu \right)f_{R_{\mu\nu}^2}(\Box)R^{\rho\sigma} = 0 \,.
\label{eq:RmnfRmnEoM}
\end{align}
One recognizes the similarity to the field equations of quadratic gravity \cite{Stelle:1977ry}, but with the constant couplings replaced by non-trivial form factors acting on the Ricci scalar $R$ and tensor $R^{\rho\sigma}$.
The full field equations generated by the effective action \labelcref{eq:EffActionwFormFactors} are then given by the sum of \labelcref{eq:RfREoM}, \labelcref{eq:RmnfRmnEoM} and the variation of $\mathcal{R}(\Box,R)$, which is non-local due to the structure of \labelcref{eq:R_operator}.

In order to generalise the classical Einstein-Hilbert term in a diffeomorphism invariant fashion we constructed the operator $\mathcal{R}(\Box,R)$ containing the non-local term $(\square + R)^{-1}$. The form factors $f_{R_{\mu\nu}^2}(\Box)$ and $f_{R^2}(\Box)$ also contain such non-local operators, c.f.\@ the analytic fits \labelcref{eq:formfactorSimpleAnalytic}, \labelcref{eq:FitFuncfRmn} and \labelcref{eq:FitFuncfR2}. This poses a non-trivial issue for numerical computations, because inverse differential operators are generically difficult to deal with and require finding Green's functions $\mathcal{G}(x,x')$ which solve 
\begin{align}
(\Box + R)\mathcal{G}(x,x') = \frac{1}{\sqrt{-g}} \ \delta(x-x')\,,
\label{eq:GreensFctDefining}
\end{align}
on general curved backgrounds \cite{Candelas:1977tt, Antonsen:1996kc}.
This task could be tackled e.g.\@ by expanding $\mathcal{G}(x,x')$ about the Feynman propagator in flat space,  $\mathcal{D}_\text{F}(x,x')$, or by means of an eigenfunction expansion using numerical techniques.
Alternatively, one could implement an iterative procedure by starting with some initial guess for the Green's function, using it to compute solutions to the field equations and inserting these back into \labelcref{eq:GreensFctDefining} in order to improve the initial guess.
This is beyond the scope of the present paper, however, and we leave a more thorough analysis to future work.
For recent studies involving non-local terms in the context of black holes see \cite{Knorr:2022kqp, Platania:2023uda}.

Here, we will instead focus on investigating the dominant IR quantum gravity effects.
Inserting the IR limits of the form factor expressions in \labelcref{eq:formfactorTaylor}, \Cref{eq:EffActionwFormFactors} results in the local infrared effective action $\Gamma_{\text{IR}}[g]$
\begin{align}\nonumber 
\Gamma_{\text{IR}}[g_{\mu\nu}]= \frac{1}{16 \pi}\int_x \sqrt{-g}
\big(
G_N^{-1} R  + g_{R_{\mu\nu}^2} R_{\mu\nu}R^{\mu\nu}
\\[1ex]
+ g_{R^2} R^2 +c_1 R_{\mu\nu}\Box R^{\mu\nu} + c_2 R \Box R
\big) \, .
\label{eq:CurvatureExpansion}
\end{align}
The field equations resulting from \labelcref{eq:CurvatureExpansion} are obtained by simply inserting the expansion of the form factors \labelcref{eq:formfactorTaylor} into the field equations \labelcref{eq:RfREoM} and \labelcref{eq:RmnfRmnEoM}, and adding Einstein's field equations resulting from the Einstein-Hilbert term $G_N^{-1}R$.

We expect the effective action \labelcref{eq:CurvatureExpansion} to capture the most relevant corrections to the classical theory, that might be detectable in future gravitational wave observatories or Event Horizon Telescope images.

%%%%%%%%%%%%%%%%%%%%%%%%%%%%%%
\subsection{Spherically symmetric solutions}
\label{sec:SphericallySymSol}

%%%%%%%%%%%%%%%%%%%%%%%%%%%%%%%%%%%%
\begin{figure*}[t]
  \centering
  \includegraphics[width=\textwidth]{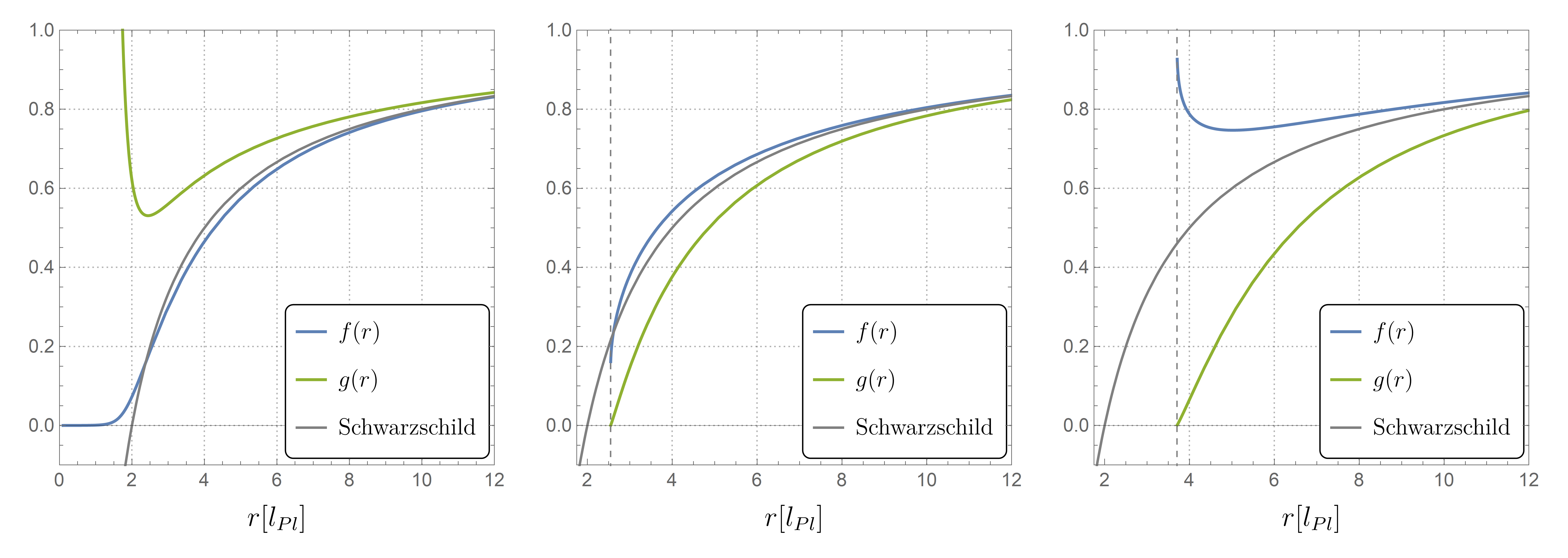}
 \caption{Examples of spherically symmetric solutions for $S_2=1$ and $M=1$. The solutions containing a zero-crossing $r_h$ are plotted only up to that radius, marked by a dashed line. The type I solution on the left has $S_0=-1$, the type II solution in the middle is found for $S_0=1$ and the type III solution shown on the right has $S_0=4$.}
  \label{fig:fgSolsRowPlot}
\end{figure*}
We restrict ourselves to static, spherically symmetric metrics.
The most general ansatz for such metrics in Schwarzschild coordinates reads
\begin{align}
ds^2 = -f(r)\text{d}t^2+\frac{1}{g(r)}\text{d}r^2 + r^2 d\Omega^2\,,
\label{eq:sphericalgeneral}
\end{align}
with two a priori independent lapse functions $f(r)$ and $g(r)$ depending on the radial coordinate $r$ only. The angular element $d\Omega^2$ is defined by
\begin{align}
d\Omega^2 = \text{d}\theta^2 + \text{sin}^2(\theta)\text{d}\varphi^2 \,.
\end{align}
The classical Schwarzschild metric is obtained for ${f(r)=g(r)=1-2M/r}$, where $M$ corresponds to the mass of the black hole that would be measured by an observer at infinity.

Our strategy for solving the field equations for $f(r)$ and $g(r)$ is as follows: We assume that at very large radii, $r_{\rm in}\gg1$, where we initialise our numerical integration, the dominant contribution is given by the terms ${R+R^2+R_{\mu\nu}^2}$. Then, we make use of the well-known weak-field limit of classical quadratic gravity, first studied in \cite{Stelle:1977ry}, in order to set the initial conditions for the metric and its derivatives. The weak-field solution is given by
\begin{align} \nonumber
f(r\gg1)&= 1 -\frac{2M}{r} + S_0 \frac{e^{-m_0 r}}{r} + S_2 \frac{e^{-m_2 r}}{r}\,, \\[1ex] \nonumber
g(r\gg1)&=1 -\frac{2 M}{r} - S_0 \frac{e^{-m_0 r}}{r}(1+m_0 r) \\[1ex]
&\phantom{=1 -\frac{2 M}{r})}+ \frac{1}{2} S_2 \frac{e^{-m_2 r}}{r}(1+m_2 r)\,,
\label{eq:WeakFieldLimit}
\end{align}
where the two free parameters $S_0$ and $S_2$ determine the strength of the exponentially decaying Yukawa corrections.
The masses $m_0$ and $m_2$ of the spin-0 and spin-2 modes, respectively, are related to the couplings of the different curvature terms. The resulting numerical values within our setup are given in \labelcref{eq:Spin0Spin2Masses}.

We choose to initialise our integration at the radius $r_{\rm in}=20$, which is located in a regime far enough from the origin such that the curvature is still small, but the exponential corrections in the initial conditions \labelcref{eq:WeakFieldLimit} are still sizeable enough to be captured by the numerics.
Inserting the general ansatz \labelcref{eq:sphericalgeneral} for the metric into the field equations derived from \labelcref{eq:CurvatureExpansion} leads to equations of motion containing up to 6 derivatives in the radial coordinate $r$.
The resulting fully non-linear system of coupled, sixth order differential equations for the functions $f(r)$ and $g(r)$ is then integrated numerically towards $r\rightarrow 0$ for different initial values $(M,S_0,S_2)$.
\\

For the initial conditions $(S_0,S_2)=(0,0)$ with $M=1$ we recover the classical Schwarzschild metric, as expected. The event horizon at $r_S=2$ is defined by the simultaneous zero-crossing ${f(r_S)=g(r_S)=0}$.
When the geometry contains a radius $r_h$ where $f(r_h)=0$ or $g(r_h)=0$, one cannot straightforwardly continue the integration towards $r=0$.
In studies of spherically symmetric solutions of quadratic gravity  \cite{Lu:2015cqa, Lu:2015psa, Holdom:2016nek, Podolsky:2018pfe, Goldstein:2017rxn, Bonanno:2019rsq, Podolsky:2019gro, Saueressig:2021wam, Daas:2022iid, Silveravalle:2022wij} this issue is dealt with by using an analytic expansion of the metric near the horizon in combination with a shooting method for the numerical integration.
However, as we are considering an IR expansion of the form factors, the solutions will not be reliable in the high curvature regime below the horizon and thus we refrain from improving the numerical strategy in this regime.

Subsequently, we consider $M=1$ and non-zero initial values for $(S_0,S_2)$.
Firstly, we observe that the behaviour of a solution is essentially determined by the value of $S_0$.
This is because for our values of the couplings $m_2>m_0$, see \labelcref{eq:Spin0Spin2Masses}, and therefore the terms $\propto S_2$ in the initial conditions \labelcref{eq:WeakFieldLimit} are suppressed. For this reason we restrict our discussion on the dependence of the solutions on $S_0$ at fixed $S_2$.
Essentially, this means that we focus on the effect of the lightest additional mode introduced by the higher curvature operators.

Depending on the value of $S_0$, the solutions can be classified into three types. One example of each type is shown in \Cref{fig:fgSolsRowPlot}.
As we currently do not have analytical expressions for these geometries at hand, we will restrict ourselves to describing some key distinguishing features of the three new solution types in the following. A more thorough geometrical analysis and interpretation will require more refined numerics and a combination with analytical methods.

Negative values of $S_0$ result in type I solutions. These are horizon-less geometries where the metric remains regular at the origin. Here, $g_{00}=f(r)$ vanishes towards the origin, whereas $1/g_{rr}=g(r)$ diverges. However, the Kretschmann scalar of these solutions diverges at the origin, hinting at a naked singularity.

For positive values of $S_0$ the solutions contain a radius $r_h$ where $g(r_h)=0$. This $r_h$ coincides with the classical Schwarzschild horizon $r_S$ for $S_0=0$.
The larger $S_0$, the further $r_h$ is shifted to larger radii. The dependence of $r_h$ on $S_0$ is found to be approximately linear, ${r_h(S_0)\propto S_0}$.

When $S_0$ remains smaller than a critical value $S_{0,\text{c}}$ we find type II solutions.
In these geometries, also $f(r)$ tends to zero at $r_h$, indicating the presence of an event horizon. We therefore interpret these solutions as modified black holes. Due to the limited numerical resolution, a finite value $f(r_h)\neq 0$, which would correspond to a wormhole geometry, can not be excluded however.

For values of $S_0$ larger than $S_{0,\text{c}}$, we also find ${g(r_h)=0}$, but $f(r)$ instead grows toward $+ \infty$ at $r_h$. We refer to these exotic geometries, for which a physical interpretation is currently missing, as type III.

Comparing the three classes of solutions, one notices that for type I solutions $g(r)> f(r)$ holds, whereas for types II and III we have $f(r)>g(r)$.
Further we note that type II and III solutions can be distinguished by the sign of the derivative $f'(r)$ at $r_h$, namely $f'(r_h)>0$ for type II and $f'(r_h)<0$ for type III.

In order to map out the phase space of solutions in the $(M,S_0)$-plane we consider the temperature
\begin{align}
T=\frac{1}{4\pi} \sqrt{\vert f'(r_h) g'(r_h) \vert}\,,
\label{eq:HawkingTemp}
\end{align}
that would be measured by a stationary observer at infinity when the spacetime contains an event horizon \cite{Borissova:2022jqj}. In the case of the Schwarzschild spacetime, \labelcref{eq:HawkingTemp} reproduces the Hawking temperature ${T_H=1/(8\pi M)}$. Although the zero-crossings of our solutions at $r_h$ may not mark a physical horizon for all solutions types, \labelcref{eq:HawkingTemp} can still be used as an order parameter for the classification.

\begin{figure}
  \centering
  \includegraphics[width=\columnwidth]{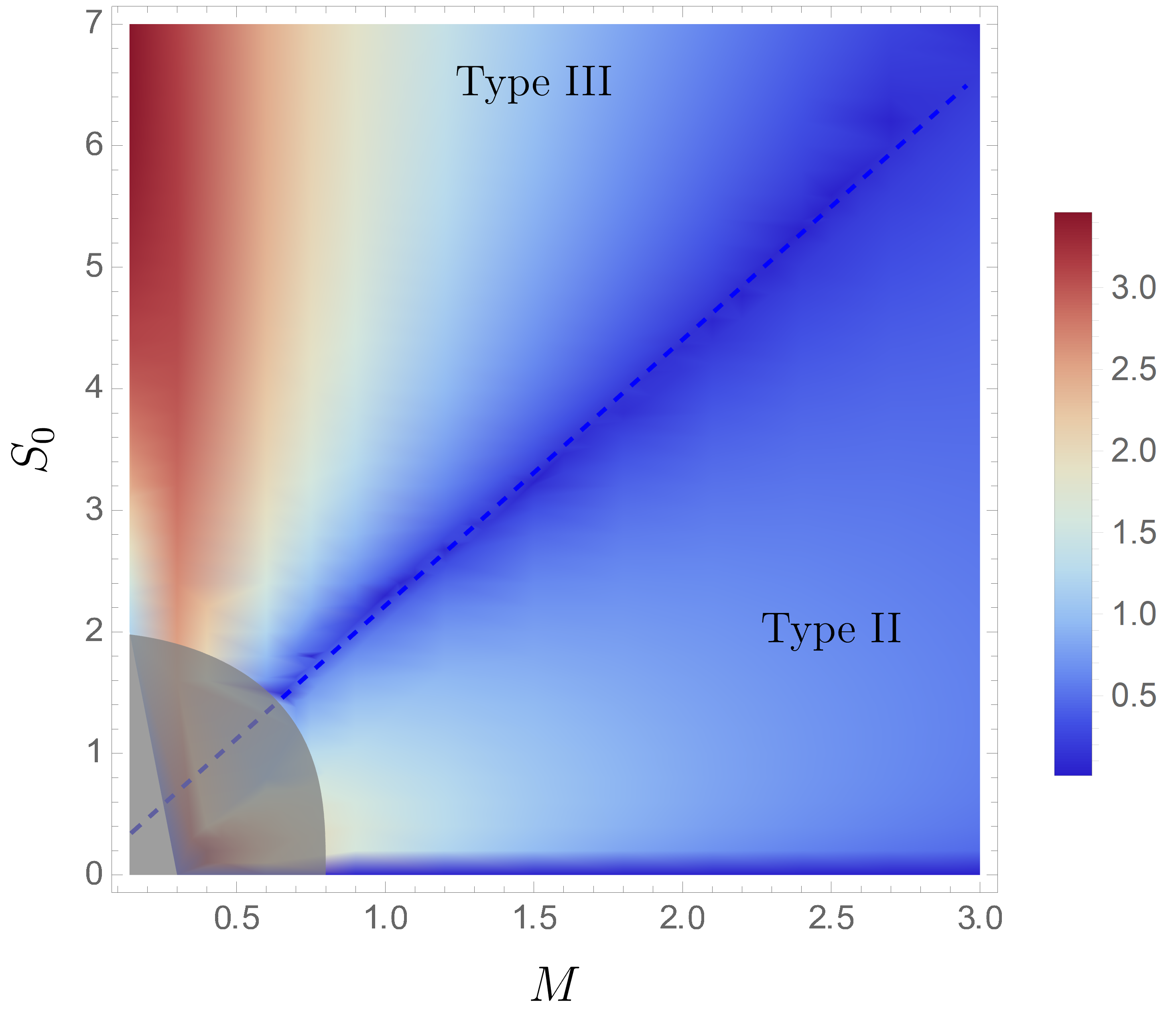}
  \caption{The phase diagram of infrared quantum gravity in the $(M,S_0)$-plane.
  The color scale indicates the value of the temperature parameter $T(M,S_0)$ as defined in \labelcref{eq:HawkingTemp}.
  The lower right corner is populated by type II solutions, and in the upper left corner type III solutions are found. A dark blue line, where $T(M,S_0)$ goes to zero, marks the phase transtion between the two solutions types.
  In the very low $M$ and $S_0$ regime in the lower left corner, $r_h\rightarrow 0$ and the numerics breaks down close to the origin.}
  \label{fig:PhaseDiag}
\end{figure}

The resulting map $T(M,S_0)$ is displayed in \Cref{fig:PhaseDiag}. The color scale indicates the value of the temperature parameter \labelcref{eq:HawkingTemp} for solutions with different values of $M$ and $S_0$. The horizontal dark blue line at $S_0=0$ indicates that the temperature of the new solutions with non-zero $S_0$ is larger than that of the corresponding Schwarzschild solution. The color gradient from left to right is consistent with the classical expectation, that black holes with small masses are hotter as $T_H\propto 1/M$.

However, we also observe a dark blue line where $T(M,S_0)$ goes to zero, marked by a dashed line in the figure. The dependence of the corresponding $S_0$ value on this line is approximately linear in $M$. For solutions on this line $f'(r_h) \rightarrow 0$, causing \labelcref{eq:HawkingTemp} to vanish. This marks a transition between the type II and III solutions, examples of which are shown in the middle and right panel of \Cref{fig:fgSolsRowPlot}, respectively. \Cref{fig:PhaseDiag} can be understood as a phase diagram of the theory: In the lower right corner of the diagram, below the phase transition line, the geometries are the modified black hole solutions of type II. In the upper left corner, where $S_0$ is large, the solutions are of the exotic type III. Type I solutions do not feature a horizon-like radius $r_h$, such that the parameter \labelcref{eq:HawkingTemp} can not be defined in these cases. This type of solutions is located at negative $S_0$ values in the phase diagram.

The structure of the phase diagram allows for interesting speculations about the Hawking evaporation and its endpoint. In particular, astrophysical black holes with masses many orders larger than the Planck mass will be located very far on the right of the phase diagram. For $\mathcal{O}(1)$ values of the parameter $S_0$ the respective geometries will be of the modified black hole type II and from the outside will resemble the Schwarzschild solution (for non-rotating black holes). Due to Hawking radiation, these black holes would gradually lose mass and move towards the left of the phase diagram (assuming $S_0$ stays constant), until they encounter the phase transition line where the temperature according to \labelcref{eq:HawkingTemp} would vanish and the evaporation would stop. For a discussion of the thermodynamic properties and stability of these possible remnants, a more detailed investigation of the limiting solutions near the phase transition is needed, however.

%%%%%%%%%%%%%%%%%%%%%%%%%%%%%%%%%
%%%%%%%%%%%%%%%%%%%%%%%%%%%%%%%%%

It would be interesting to study the relation between our numerical solutions and other non-Schwarzschild geometries that have been found in higher derivative gravity \cite{Lu:2015cqa, Lu:2015psa, Holdom:2016nek, Podolsky:2018pfe, Goldstein:2017rxn, Bonanno:2019rsq, Podolsky:2019gro, Saueressig:2021wam, Daas:2022iid, Silveravalle:2022wij}. In particular it would be interesting to investigate whether the solutions we found can be classified into the same families of solutions in terms of a Frobenius expansion near the horizon. 
For example, our type I solution in \Cref{fig:fgSolsRowPlot} resembles the type II solution plotted in \cite{Silveravalle:2022wij}, while our type II solution could be related to the non-Schwarzschild black hole or type III solution of \cite{Silveravalle:2022wij}. However, for a definitive comparison a more detailed analysis of the causal structure and physical properties of our new geometries is needed and is left for future work.

In \cite{Lu:2015psa, Daas:2022iid} it has been argued that in a generic quadratic gravity theory any asymptotically flat solution containing a horizon must have vanishing Ricci scalar $R=0$ in the region outside the horizon. For generic $g_{R^2}\neq 0$ and $g_{R_{\mu\nu}^2}\neq 0$ this seems to single out the Schwarzschild solution as the only asymptotically flat solution with a horizon (for sufficiently massive black holes). It would be interesting to investigate whether similar restrictive statements apply for our type of action \labelcref{eq:EffActionwFormFactors} with non-trivial form factors.

In conclusion, we regard it a promising sign that the inclusion of higher order curvature operators in the asymptotically safe effective action \labelcref{eq:CurvatureExpansion} allows solutions that approach the classical Schwarzschild metric in the IR regime, but show new, non-trivial behaviour at larger curvatures.
The above discussion indicates that the higher derivative corrections may alter the classical geometry significantly already outside the classical horizon.

%%%%%%%%%%%%%%%%%%%%%%%%%%%%%%%%%%%%
\section{Summary}
\label{sec:Summary}

In this work we studied the quantum effective action and black hole solutions in asymptotically safe gravity.

The key object of our investigation is the background effective action $\Gamma[\bar{g}_{\mu\nu}]$, which we studied in detail in \Cref{sec:ReconstructingGamma}.
We parametrised $\Gamma[\bar{g}]$ by generalising the classical Einstein-Hilbert term to the operator $\mathcal{R}(\Box,R)$ and by including the two form factors $f_{R_{\mu\nu}^2}(\Box)$ and $f_{R^2}(\Box)$ to capture higher order effects.
These form factors were reconstructed from the momentum-dependent data of graviton $n$-point correlation functions.

In particular, we computed the fluctuation scattering coupling $\gamma_h^{(4)}(p)$ of the four-graviton vertex $\Gamma^{(0,4)}$. It shows a classical regime in the IR, where $\gamma_h^{(4)}$ is constant, and falls off like $p^{-2}$ in the UV, compatible with AS. 
It is displayed in \Cref{fig:MomDepG4}.

The reconstructed form factors take constant values both in the IR and the UV limits and were found to contain non-local contributions, c.f.\@ \labelcref{eq:R_operator} and \labelcref{eq:formfactorSimpleAnalytic}.
The momentum dependence of $f_{R_{\mu\nu}^2}(p^2)$ and $f_{R^2}(p^2)$ is plotted in \Cref{fig:FormFactors}.
Importantly, our work presents a mapping between the scattering-couplings of gravitons and diffeomorphism invariant operators in the effective action.

In \Cref{sec:QuBHSolutions} we studied spherically symmetric solutions of the effective field equations in our theory.
Focussing on low-energy corrections to the classical Schwarzschild black hole, we expanded the form factors about vanishing $p^2$, which results in sixth order derivatives in the field equations.

Depending on the initial values for the numerical integration, our solutions can be categorized into three types, displayed in \Cref{fig:fgSolsRowPlot}. 
We observed intriguing hints that the quantum corrections may shift the location of the classical horizon and significantly alter the classical geometry already in the near-horizon regime, leading to naked singularities (type I), modified black holes (type II) or exotic type III geometries.
It is interesting that the additional degrees of freedom in the field equations allow for solutions that resemble a classical Schwarzschild black hole in the IR but show new, non-trivial features in the UV.

Promising places to look for observational signatures of these new solutions would be the shadow images of supermassive black holes \cite{Held:2019xde, Eichhorn:2022bgu, Daas:2022iid} or the ringdown of the gravitational wave signal after a compact binary merger \cite{Silva:2022srr}. Although, based on the exponential suppression of the corrections in the weak field limit \labelcref{eq:WeakFieldLimit}, one would naively expect the modifications to the classical solution to be small for astrophysical masses $M\gg1$, one may still be able to constrain the possible values of $S_{0}$ and $S_2$. Furthermore, the effects may be significantly enhanced when considering rotating black holes \cite{Eichhorn:2022bbn}.

In summary, we interpret our results as a promising and important step towards connecting the fundamental asymptotically safe theory of quantum gravity to potentially observable consequences for black hole spacetimes.

%%%%%%%%%%%%%%%%%%%%%%%%%%%%%%%%%%%%%%%%%%%%%%%%%%%
\acknowledgments
We thank A.~Bonanno, T.~Denz, A.~Held, A.~Eichhorn and M.~Reichert for discussions. This work is funded by the Deutsche Forschungsgemeinschaft (DFG, German Research Foundation) under Germany’s Excellence Strategy EXC 2181/1 - 390900948 (the Heidelberg STRUCTURES Excellence Cluster) and the Collaborative Research Centre SFB 1225 - 273811115 (ISOQUANT).
J.T. is supported by the DFG under the Emmy-Noether program grant no. DO 2574/1-1, project number 496592360.

%%%%%%%%%%%%%%%%%%%%%%%%%%%%%%%%%%%%%%%%%%%%%%%%%%%
\appendix

%%%%%%%%%%%%%%%%%%%%%%
\section{Gauge fixing}
\label{sec:GaugeFixing}
We employ a standard linear gauge fixing of the form
\begin{align}
S_{gf}[\bar{g},h] = \frac{1}{2\alpha}\int_x \sqrt{\bar{g}} \ \bar{g}^{\mu\nu}F_\mu F_\nu\,,
\label{eq:GaugeFixingAction}
\end{align}
where we choose a de-Donder type gauge-fixing condition
\begin{align}
F_\mu[\bar{g},h]=\bar{\nabla}^\nu h_{\mu\nu} - \frac{1+\beta}{4}\bar{\nabla}_\mu h^\nu_{\phantom{\nu}\nu} \,.
\end{align}
For the gauge parameters $\alpha$ and $\beta$ we choose the harmonic gauge $\beta=1$ and the Landau limit $\alpha\rightarrow 0$. We enforce this condition in the path integral with the Faddeev-Popov method and introduce the ghost fields $c_\mu$ and $\bar{c}_\mu$. The ghost action reads
\begin{align}
S_{gh}[\bar{g},h,c,\bar{c}] = \int_x \sqrt{\bar{g}} \ \bar{c}^\mu \mathcal{M}_{\mu\nu} c^\nu \,,
\label{eq:GhostAction}
\end{align}
and the Faddeev-Popov operator is defined by
\begin{align}
\mathcal{M}_{\mu\nu} =\bar{\nabla}^\rho \left(g_{\mu\nu}\nabla_\rho + g_{\rho\nu}\nabla_\mu \right) - \frac{1+\beta}{2}\bar{g}^{\sigma\rho}\bar{\nabla}_\mu g_{\nu\sigma}\nabla_\rho \,.
\end{align}
%

%%%%%%%%%%%%%%%%%%%%%%%%%%%%%%%%%%%%
\section{Vertex expansion}
\label{app:VertexExpansion}

Written explicitly, the vertex expansion of the effective action $\Gamma_k[\bar{g},\phi]$
shown in \Cref{fig:VertexExpansion} reads
\begin{align}
\Gamma_k[\bar{g},\phi] = \displaystyle\sum_{n=0} ^{\infty} \frac{1}{n!} \Gamma_k^{(0,\phi_{i_1}...\phi_{i_n})}[\bar{g},0] \prod_{l=1}^n \phi_{i_l} \,,
\label{eq:VertexExpansion}
\end{align}
where summation over indices $i_l$ that appear twice is understood. In \labelcref{eq:VertexExpansion} we summarised the fluctuating graviton $h_{\mu\nu}$ and the ghost and anti-ghost fields $c_\mu$ and $\bar{c}_\mu$ in the fluctuation superfield 
\begin{align}
\phi = (h,\bar{c},c)\,.
\label{eq:SuperField}
\end{align}
In this work we will make a similar ansatz for the $n$-point functions $\Gamma^{(n)}$  as has already been employed in \cite{Christiansen:2014raa, Christiansen:2015rva, Denz:2016qks, Christiansen:2016sjn,Meibohm:2015twa}. The considerations explained in \Cref{sec:VertExp} amount to the parametrisation of the $n$-point functions
\begin{align}
\Gamma_k ^{(0,\phi_1...\phi_n)}(\boldsymbol{p}) = \left( \prod_{i=1}^n Z_{\phi_i}^{\frac{1}{2}}(p_i^2)\right) \gamma_h^{(n)}(\boldsymbol{p})\mathcal{T}^{(\phi_1...\phi_n)}(\boldsymbol{p};\Lambda_n) \,,
\label{eq:nPtfunction}
\end{align}
where we approximate the complete set of tensor structures $\mathcal{T}^{(\phi_1...\phi_n)}$ by the ones generated by the classical Einstein-Hilbert action \labelcref{eq:EinsteinHilbertAction}. The cosmological constant is replaced by $\Lambda\rightarrow\Lambda_n$, capturing the momentum-independent part of the correlation functions.
The tensor structures $\mathcal{T}^{(\phi_1...\phi_n)}$ are then given by
\begin{align}
\mathcal{T}_\textrm{EH}^{(\phi_1...\phi_n)}(\boldsymbol{p};\Lambda_n)= G_N S_{\text{EH}}^{(\phi_1...\phi_n)}(\boldsymbol{p}; \Lambda \rightarrow \Lambda_n) \,.
\label{eq:EHtensorstruct}
\end{align}
Note that even though we refrain from explicitly including tensor structures derived from higher order curvature invariants, they will typically still be generated by the RG-flow and their contribution is carried by the momentum dependence of the $n^{\text{th}}$-order dressings $\gamma_h^{(n)}(\boldsymbol{p})$.

In this work we consider a maximally symmetric ${(n-1)}$-simplex configuration for all vertices, meaning that $\vert p_i \vert = p$ for all $n$ momenta.
In this configuration the dressings depend only on the average momentum $p^2$ flowing through the vertex.
Equal angles between all momenta imply that the scalar product between two momenta $i,j$ is given by ${p_i\cdot~p_j=(n\delta_{ij}-1)/(n-1)\cdot~p^2}$.

The dimensionless flow of the graviton $n$-point functions is defined as
\begin{align}
\text{Flow}^{(n)}(\hat{p}) \coloneqq \frac{\partial_t \Gamma^{(n)}(\hat{p})}{Z_h^{n/2}k^{4-n}} \,,
\label{eq:Flow(n)_G}
\end{align}
where $\hat{p}$ is the dimensionless momentum $\hat{p}=p/k$. The dimensionless counterparts of the couplings are given by
\begin{align}
g_n\coloneqq \left( \gamma_h^{(n)} \right)^{\frac{2}{n-2}} k^2, \quad \lambda_n \coloneqq \Lambda_n k^{-2}, \quad \mu_h \coloneqq -2 \lambda_2 \,.
\label{eq:dimlesscouplings}
\end{align}
Their flows vanish in the scale invariant regime at the fixed point.
$\mu_h$ represents the momentum independent part of the graviton propagator and is referred to as the graviton mass parameter.

In order to technically facilitate the inversion of the two-point function, which is in general a complicated rank-4 tensor, we employ a Stelle decomposition of the full graviton field \cite{Stelle:1976gc}.

The derivation of symbolic flow equations, the computation of the tensor structures and the contractions from the projection scheme are done using specialised computer algebra programs.
Numerical integrations in this work have been carried out with Mathematica's native NDSolve method \cite{NDSolve}.
More information on the computational details can be found in \cite{Denz:2016qks}.

%%%%%%%%%%%%%%%%%%%%%%%%%%%%%%%%%%%%
\section{Projection scheme}
\label{app:ProjectionScheme}

As we are expanding about a flat background we can work in momentum space. Then, the tensor structures will be combinations of momenta and background metrics with index structures reflecting the correct symmetries. We project on the transverse-traceless mode, which is numerically dominant as well as gauge-independent and hence expected to capture the most relevant physical information \cite{Denz:2016qks,Christiansen:2016sjn}.
The tt-projection operator $\Pi_{tt}$ for the transverse-traceless mode is defined in terms of the transversal projector $\Pi_{\bot}$ and reads
\begin{subequations} \label{eq:TT_projector} 
\begin{align}
\Pi_{\textrm{tt}}^{\mu\nu\alpha\beta} &\coloneqq \frac{1}{2} \left(\Pi_\bot^{\mu\alpha}\Pi_\bot^{\nu\beta} + \Pi_\bot^{\mu\beta}\Pi_\bot^{\nu\alpha} \right) - \frac{1}{3} \Pi_\bot^{\mu\nu}\Pi_\bot^{\alpha\beta}\,,
\end{align}
with the transverse projection operator 
\begin{align}
\Pi_\bot^{\mu\nu} &\coloneqq \delta^{\mu\nu} - \frac{p^{\mu}p^{\nu}}{p^2} \,.
\label{eq:transverse_projector}
\end{align}
\end{subequations}
The completely transverse-traceless part of the curvature tensor structure of the classical action in the flat background is given by contracting all legs of ${{\cal T}_\textrm{EH}^{(n)}(\boldsymbol{p},\Lambda_n= 0)}$ with the tt-projection operator $\Pi_\textrm{tt}$ to wit 
\begin{align} 
	\mathcal{T}_{R,\textrm{tt}}^{(n)}(\boldsymbol{p})= \prod_{i=1}^n \Pi_\textrm{tt}(p_i) {\cal T}_\textrm{EH}^{(n)}(\boldsymbol{p}, 0) \, .
	\label{eq:TTPartSEHn}
\end{align}	 
In our approximation \labelcref{eq:NIbargh} for the Nielsen identity, the full $n$-point functions take the form 
\begin{align}\nonumber 
		\Gamma^{(0,n)}(\boldsymbol{p})  =&\, \prod_{i=1}^{n}\sqrt{Z_h(p_i) \,G(p_i)  }\, \gamma_g^{(n)}(\boldsymbol{p}) \,\mathcal{T}_{R,\textrm{tt}}^{(n)}(\boldsymbol{p})+\cdots\,, \\[1ex]
	\Gamma^{(n)}(\boldsymbol{p})  =&\, \gamma_g^{(n)}(\boldsymbol{p}) \mathcal{T}_{R,\textrm{tt}}^{(n)}(\boldsymbol{p})+\cdots \,, 
	\label{eq:ttGamman}
\end{align}
where the $\cdots$ indicate a sum over the remaining tensors structures ${\cal T}^{(n)}_i$ with respective fully momentum-dependent coefficients in a complete basis expansion with the basis $\{{\cal T}^{(n)}_i\}$ with $i=( (R,\textrm{tt}), \cdots)$. 

We now assume an orthogonal basis and project on the coefficients of the tt-part of the $n$-point functions. For the fluctuation $n$-point functions we arrive at 
\begin{align}\nonumber 
 \gamma_g^{(n)}(\boldsymbol{p}) =&\,\prod_{i=1}^{n}\frac{1}{\sqrt{Z_h(p_i) \,G(p_i)  }}\,\\[1ex] 
 &\,\times \frac{  
		\Gamma_{\mu_1\nu_1 \cdots \mu_n \nu_n}^{(0,n)} { \mathcal{T}_{R,\textrm{tt}}^{(n)\, \mu_1\nu_1 \cdots \mu_n \nu_n}}}{{\mathcal{T}_{R,\textrm{tt}}^{(n)}}_{\mu_1\nu_1 \cdots \mu_n \nu_n} {\mathcal{T}_{R,\textrm{tt}}^{(n) \, \mu_1\nu_1 \cdots \mu_n \nu_n}}} \,,
\label{eq:gammanFlucApp}
\end{align}
while  for the background $n$-point functions we get 
\begin{align}
\gamma_g^{(n)}(\boldsymbol{p})=\frac{  
	\Gamma_{\mu_1\nu_1 \cdots \mu_n \nu_n}^{(n)} {\mathcal{T}_{R,\textrm{tt}}^{(n) \, \mu_1\nu_1 \cdots \mu_n \nu_n}}}{{\mathcal{T}_{R,\textrm{tt}}^{(n)}}_{\mu_1\nu_1 \cdots \mu_n \nu_n} {\mathcal{T}_{R,\textrm{tt}}^{(n) \, \mu_1\nu_1 \cdots \mu_n \nu_n}}} \,.
\label{eq:gammanBackApp}
\end{align} 
Written more explicitly, the projection of the fluctuation graviton $n$-point function \labelcref{eq:nPtfunction} results in
\begin{align}
\Gamma_{G}^{(0,n)}(p^2) & \coloneqq \mathcal{T}_{R,\textrm{tt}}^{(n)}/p^2 \circ \Gamma_k^{(0,n)} \nonumber \\[1ex]
&= Z_h^{\frac{n}{2}} g_n^{\frac{n}{2}-1}k^{2-n} \left(C^{(n)}_{\Lambda_n}\lambda_n k^2 + C_{R}^{(n)}p^2\right) \,,
\label{eq:ProjectednPtfunction}
\end{align}
where $\circ$ denotes the pairwise contraction of Lorentz indices and we divide $\mathcal{T}_{R,\textrm{tt}}^{(n)}$ by $p^2$ to ensure that the projector is dimensionless. The subscript $G$ is used from now to indicate this projection. The coefficients $C_{\cal{O}}^{(n)}$ quantify the respective overlap and are given below.

This projection scheme also allows us to compute the overlap of tensor structures generated by higher curvature operators like $R^2$ and $R_{\mu\nu}^2$ with our approximation of $\Gamma_{G}^{(0,n)}$. This is done by taking the contraction
\begin{align}
\mathcal{T}_{R,\textrm{tt}}^{(n)}/p^2\circ S^{(n)}_{R^2+R_{\mu\nu}^2} = C_{R^2}^{(n)} g_{R^2} p^4 + C_{R_{\mu\nu}^2}^{(n)} g_{R_{\mu\nu}^2} p^4 \,,
\label{eq:projection_scheme}
\end{align}
with $S_{R^2+R_{\mu\nu}^2}=\int_x \sqrt{g}\left(  g_{R^2} R^2 + g_{R_{\mu\nu}^2} R_{\mu\nu}^2 \right) $.

In \cite{Denz:2016qks} these overlaps have been computed.
The explicit values of the numerical constants are
\begin{align}
C_{\Lambda_3}^{(3)} &= -\frac{9}{4096 \pi^2}\,,  & C_{\Lambda_4}^{(4)} &= \frac{222485}{60466176\pi^2} \,,
\nonumber \\[1ex]
C_{R}^{(3)} &= \frac{171}{32768\pi^2}\,,  & C_{R}^{(4)} &=\frac{6815761}{544195584\pi^2} \,,
\nonumber \\[1ex]
C_{R_{\mu\nu}^2}^{(3)}& = -\frac{405}{32768 \pi^2}\,, & C_{R_{\mu\nu}^2}^{(4)} &=-\frac{3676621}{51018336 \pi^2} \,,
\nonumber \\[1ex]
C_{R^2}^{(3)}&=0\,, & C_{R^2}^{(4)} &=-\frac{96203921}{1632586752\pi^2} \,.
\label{eq:projection_coefficients}
\end{align}
%

%%%%%%%%%%%%%%%%%%%%%%%%%%%%%%%%%%%%
\section{Regulator}
\label{sec:Regulator}

Here we specify the regulator function $\mathcal{R}_k(p^2)$.
We will adopt the common choice to take the regulator as the product of a dimensionless shape function $r(\hat{p}^2)$, which depends on the dimensionless momentum $\hat{p}=p/k$, and the tensor structures of the kinetic part of the respective two-point function. For the graviton this reads 
\begin{align}\nonumber 
\mathcal{R}_k(p^2) &= \Gamma_k^{(0,hh)}(p^2)\bigg\vert_{\mu_h=0} r(\hat{p}^2)\\[1ex]
&= Z_h(p^2)\mathcal{T}^{(2)}(p^2;0)r(\hat{p}^2)\,,
\end{align}
and for the ghost $Z_c$ and $\mathcal{T}^{(\bar{c}c)}$ are inserted, respectively. 
For the shape function we choose the Litim regulator \cite{Litim:2001up}
\begin{align}
r(\hat{p}^2) = \left(\frac{1}{\hat{p}^2} - 1\right) \Theta (1-\hat{p}^2)\,,
\end{align}
which has the advantage that the loop integrals at vanishing external momentum $p=0$ and that of the tadpole can be solved analytically.

%%%%%%%%%%%%%%%%%%%%%%%%%%%%%%%%%%%%
\section{Flow of the four-point function}
\label{sec:FlowofFourPtFct}
In order to compute the flow of the four-graviton coupling $\gamma_h^{(4)}(p^2)$ we apply a scale derivative $\partial_t$ to the projected $n$-point function \labelcref{eq:ProjectednPtfunction}, evaluate the resulting equation bilocally at two momenta $p_1^2 = p^2$ and $p_2^2=0$ and subtract the results.
In this way the flow of the momentum independent $\lambda_4$ drops out.
We further make an approximation which relies on $\lambda_4$ remaining small (which we guarantee in the truncation we will later employ) and assumes a typical shape of the RG-trajectories $g_{4,k}(p)$.
This results in the beta-function for $g_4(p)$
\begin{align}
\nonumber \frac{\dot{g}_4(p^2)}{g_4(p^2)} = & \ 2\left(1 + \eta_h(p^2)\right)
 + 2 C_4 \frac{k^2}{p^2}(\eta_h(p^2) - \eta_h(0))\lambda_4\\[1ex]
 &+ \frac{1}{C_{R}^{(4)}}\frac{k^2}{p^2}\left(\frac{\text{Flow}^{(4)}_{G}(p^2)}{g_4(p^2)} - \frac{\text{Flow}^{(4)}_{G}(0)}{g_4(0)}\right)\,,
\label{eq:betag4}
\end{align}
where $\dot{g}_4 \coloneqq \partial_t g_4$ and we introduced the ratio ${C_4\coloneqq C^{(4)}_{\Lambda_4}/C_{R}^{(4)}}$. Note that \labelcref{eq:betag4} corresponds to equation (E5) in \cite{Denz:2016qks} for a generic momentum dependence. By construction the wavefunction $Z_{\phi_i}$ itself does not enter the flow equations, but only the anomalous dimension defined by 
\begin{align}
\eta_{\phi_i} (p^2) \coloneqq -\partial_t \ln (Z_{\phi_i}(p^2)) \,.
\label{eq:AnomalousDim}
\end{align}
\Cref{eq:betag4} involves the flow of the four-point function $\text{Flow}^{(4)}_G$, which is computed in terms of the diagrams given in \Cref{fig:Flow4PtFct}. The structure of the Wetterich equation \labelcref{eq:WetterichEq} entails that the flow of each $n$-point function depends on the ($n+1$)- and ($n+2$)-point functions, c.f.\@ \Cref{fig:Flow4PtFct}.
In order to close this infinite tower of coupled equations we make an ansatz for the highest order couplings, whose flows we do not compute.
In the present work we choose the identifications 
\begin{align}
g_6=g_5=g_4 \quad \text{and} \quad \lambda_6=\lambda_5=\lambda_3 \,.
\label{eq:identificationscheme}
\end{align}
This identification has proven to be the most numerically stable and is also used in \cite{Denz:2016qks}, which we build upon.

As we are working at the momentum symmetric point, we do not feed back the dependence of $g_n$ on the loop-momentum $q$ in the diagrams.
This approximation has been found to hold quantitatively in previous studies \cite{Christiansen:2015rva,Denz:2016qks} and technically simplifies the computation considerably.
The momentum dependence of the graviton anomalous dimension is approximated by $\eta_h(q)\approx \eta_h(k)$, which is expected to capture the leading contribution because the regulator derivative $\dot{\mathcal{R}}_k$ is peaked at $p\approx k$.
For simplicity, we will further neglect the momentum dependence of the ghost and set $\eta_c = 0$.
The remaining dependence on the loop-momentum $q$ carried by the tensor structures $\mathcal{T}^{(\phi_1...\phi_n)}(p^2,q;\Lambda_n)$ can be integrated numerically to obtain $\text{Flow}^{(4)}_G(p^2)$.
We use existing data $\eta_{h,k}(p)$ and $g_{3,k}(p)$ of trajectories that have been computed in previous work \cite{Bonanno:2021squ} to feed back these momentum dependencies into the flow of $g_4$.
Using the Newton constant $G_N = M_{\text{Pl}}^{-2}$, the Planck scale is set dynamically by demanding $\gamma_h^{(3)}(0)=G_N\equiv 1$ at $k=0$ and from now on we measure $k$ in units of $M_{\text{Pl}}$.

In \cite{Denz:2016qks} it was found that the $\lambda_n$ only give subleading contributions to the flow and we impose $\Lambda_n = 0$ at $k=0$, which resembles a vanishing cosmological constant in the IR.
We therefore choose to set $\mu_h$ and $\lambda_n$ to constant values.
In this way we feed back part of their contribution without having to solve their flows separately.
To select reasonable values for $\mu_h$, $\lambda_3$ and $\lambda_4$ we apply the following set of criteria based on results from the analysis in \cite{Denz:2016qks}:
%We want to select values such that our system reflects:
the ordering of relative sizes $\vert\mu_h^*|>\lambda_3^*>\lambda_4^*$ at the fixed point,
small values $\lambda_3,\lambda_4 < 1$, the relation
$g_4^*<g_3^*$ of fixed point values,
a fixed point with two relevant directions and
real fixed point values $g_4^*\in \mathbb{R}$.
In order to find values for $\mu_h$ and $\lambda_n$ that fulfil these requirements we investigate the fixed point structure of the flows for different choices.
We choose to tune $\mu_h$ and $\lambda_n$ so that the ratio $g_3^*/g_4^* \approx 1.46$ matches the relative values of $g_3$ and $g_4$ at the UV fixed point of the full system. This has the advantage that it reflects the relative strength of the two couplings, even though we use data for $g_3$ from a different truncation.
This is achieved for the values
\begin{align}
\mu_h=-0.26, \qquad \lambda_3 = 0.145 \quad \text{and} \quad \lambda_4 = 0.025 \,,
\label{eq:truncationlambdas}
\end{align}
which we from now on fix. After applying our identification scheme \labelcref{eq:identificationscheme} for the higher couplings and the constant values \labelcref{eq:truncationlambdas} for $\lambda_3$, $\lambda_4$ and $\mu_h$, the Flow schematically reads
\begin{align}
\text{Flow}^{(4)}_{G}(\hat{p})=&
\frac{g_3^2}{g_4}\left( c_1+c_2 \eta_h\right)+\sqrt{g_3 g_4}\left(c_3 + c_4 \eta_h \right) \nonumber \\[1ex]
&+g_3 \left(c_5+ c_6\eta_h\right)+g_4\left(c_7 + c_8 \eta_h\right) \,,
\end{align}
where the $c_i$ are functions of the momentum $\hat{p}$.

In this setup we find a UV fixed point at the values
\begin{align}
(g_3^*,g_4^*)=(2.15,1.48) \,,
\end{align}
with critical exponents ${(\theta_1,\theta_2)=(-2,-0.85)}$ indicating two relevant directions. We can now integrate the differential equation $\partial_t g_4$ to obtain trajectories $g_{4,k}(p)$. First, we solve the analytic flow at $p=0$ found in \cite{Denz:2016qks} with the initial condition chosen such that $g_3(0)=g_4(0)$ as $k\rightarrow 0$. At $p\neq 0$ we use $\hat{p}=p/k$ to distinguish the dependence on the external momentum $p$ and the cutoff-scale $k$ and integrate the beta-function \labelcref{eq:betag4} for a grid of $p$-values. As initial condition we set a value slightly below the fixed point $g_4^*$ at large $k$ and integrate the equation towards $k \rightarrow 0$. This results in one RG-trajectory $g_{4,k}(p)$ for each $p$-value. 
\begin{figure}[t]
  \centering
  \includegraphics[width=\columnwidth]{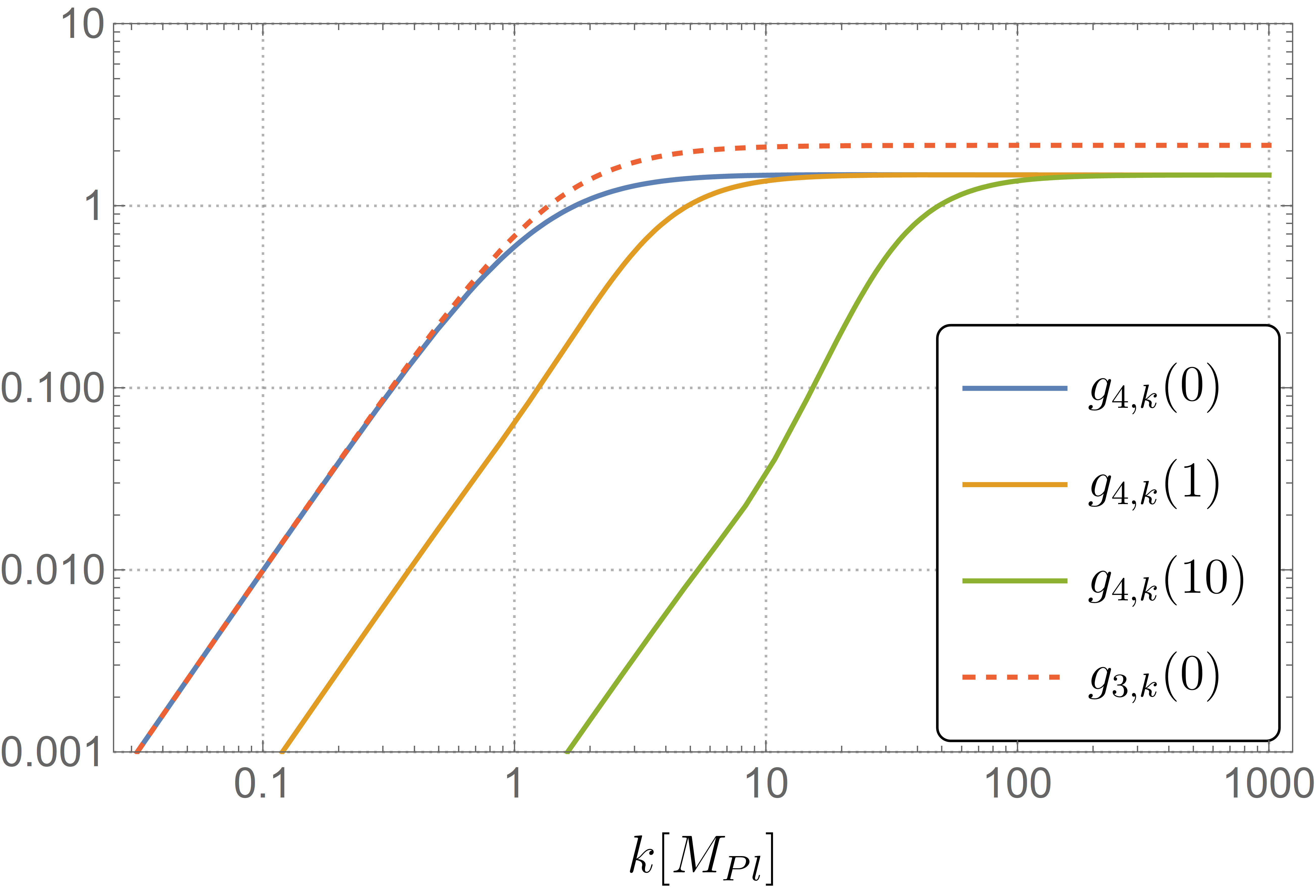}
  \caption{Examples of RG-trajectories $g_{4,k}(p)$ for different values of $p/M_{\text{Pl}}$. All scale canonically $\propto k^{2}$ below the Planck scale at $k=1$ and approach the fixed point value $g_4^*\approx 1.48$ in the UV. For reference $g_{3,k}(0)$ is also shown.}
  \label{fig:g3g4trajectories}
\end{figure}
Some example trajectories are plotted in \Cref{fig:g3g4trajectories}.
One can observe the canonical scaling $\propto k^2$ below the Planck scale at $k=1$. In the regime $k\gg 1$ all trajectories approach the UV fixed point $g_4^*$.
For larger external momentum $p$ the asymptotically safe regime is reached at larger $k$, approximately when $k \gtrsim 10 p$.

To obtain $\gamma_h^{(4)}(p)$ in the physical limit of vanishing cutoff $k\rightarrow 0$, we use \labelcref{eq:dimlesscouplings} to transform back to the dimensionful coupling by multiplying the trajectories $g_{4,k}$ with $k^{-2}$.
Upon doing this, the trajectories become constant in the IR and scale like $k^{-2}$ in the UV. Hence, we can simply read out the constant values in the IR, i.e.\@ $k\rightarrow 0$, for each value of $p$.
The resulting momentum dependent four-graviton coupling $\gamma_h^{(4)}(p)$ at vanishing cutoff is displayed in \Cref{fig:MomDepG4}.

%%%%%%%%%%%%%%%%%%%%%%%%%%%%%%%%%%
\section{The generalized Einstein-Hilbert term}
\label{sec:GenEHTerm}

The dimensionless constant $\bar{\gamma}_3$ appearing in the definition \labelcref{eq:R_operator} of $\mathcal{R}$ is defined by
\begin{align}
\bar{\gamma}_3 = \lim \limits_{p^2 \to \infty}\left(\frac{\gamma_g^{(3)}(p^2)}{p^2}\right)\,,
\label{eq:gammabar3}
\end{align}
and its numerical value computed from our data is
\begin{align}
\bar{\gamma}_3 \approx 0.054 \,.
\label{eq:gamma3UVlimit}
\end{align}
As mentioned above, $\mathcal{R}$ contributes a term ${C_{\mathcal{R}}^{(n)}g_{\mathcal{R}} p^2}$ to the flow of the $n$-point functions $\bar{\Gamma}_G^{(n)}(p)$.
In particular, for $n=3,4$ this contribution stems from the structure
\begin{align}
R^{(n)}\left( \frac{\gamma_g^{(3)} -\bar{\gamma}_3 \Delta}{\Delta + R}R \right)\Big\vert_{\Delta\rightarrow 0,R\rightarrow 0} \,.
\label{eq:p2contributionR}
\end{align}
Other derivative structures vanish due to our expansion about $R=0$ and the tt-projection.
Because the field derivatives do not hit the term in brackets there is no associated momentum flow and $\Delta\rightarrow 0$ is taken before the evaluation at $h=0$ on the flat background with $R=0$.

This implies that the constant $g_{\mathcal{R}}$ which appears in the projected $\bar{\Gamma}_G^{(n)}(p)$ is given by ${g_{\mathcal{R}} =\gamma_g^{(3)}(0) = G_N^{-1}}$.

From \labelcref{eq:p2contributionR} we further see that the $p^2$ contribution to the three- and four-point function generated by $\mathcal{R}$ has the same overlap with our parametrization as the tensor structures generated by $R$, i.e.\@ ${C_{\mathcal{R}} = C_{R}}$ for $n=3,4$.

%%%%%%%%%%%%%%%%%%%%%%%%%%%
\section{Numerical values of couplings}
\label{sec:NumericalValues}

The numerical values of the low energy couplings $g_{R_{\mu\nu}^2}$ and $g_{R^2}$ can be computed from our numerical data for $\gamma_g^{(3)}(p)$ and $\gamma_g^{(4)}(p)$ according to \labelcref{eq:gRmunu+gR2}.
Our construction results in the values
\begin{align}
g_{R_{\mu\nu}^2} \approx -0.40 \,, \qquad g_{R^2} \approx 1.9 \,.
\label{eq:IRcouplingsvalues}
\end{align}
We note that $g_{R_{\mu\nu}^2}$ in our setup turns out to be negative, whereas $g_{R^2}$ is positive.
The observational constraints on the couplings of higher curvature terms are still weak \cite{Berry:2011pb}, such that our values are easily within the allowed region.
In the UV regime the form factors approach the values
\begin{align}
\tilde{g}_{R_{\mu\nu}^2} \approx -0.023 \,, \qquad \tilde{g}_{R^2} \approx 0.028 \,.
\label{eq:UVcouplingsvalues}
\end{align}
The coefficients of the first order Taylor expansion  \labelcref{eq:formfactorTaylor} take the numerical values
\begin{align}
c_1 = 344.09 \,, \quad c_2 = -136.75 \,,
\label{eq:TaylorCoefficients}
\end{align}
in Planck units, and
\begin{align}
c_3 = \frac{C_{R}^{(3)}}{C_{R_{\mu\nu}^2}^{(3)}} \,, \qquad
c_4 = \frac{C_{R}^{(4)}}{C_{R^2}^{(4)}} \,, \qquad
c_5 = \frac{C_{R_{\mu\nu}^2}^{(4)}}{C_{R^2}^{(4)}} \, c_3 \,.
\label{eq:IRcoefficients}
\end{align}
Since the seminal work \cite{Stelle:1977ry} it is known that the spectrum of quadratic gravity, i.e.\@ of an action containing $R+R^2+R_{\mu\nu}^2$ terms, in addition to the massless graviton contains a massive spin-two ghost and a massive spin-zero excitation with masses $m_2$ and $m_0$, respectively. They are given by
\begin{align}
m_2^{\phantom{2}2} & = \frac{-G_N^{-1}}{g_{R_{\mu\nu}^2}} \approx 2.5 \, M_{\text{pl}}^2 \,, \nonumber \\
m_0^{\phantom{0}2} & = \frac{G_N^{-1}}{6 g_{R^2} + 2g_{R_{\mu\nu}^2}} \approx 0.095 \, M_{\text{pl}}^2 \,,
\label{eq:Spin0Spin2Masses}
\end{align}
using the values \labelcref{eq:IRcouplingsvalues} of the couplings in our setup. We note that, due to the signs and relative sizes of the couplings in \labelcref{eq:IRcouplingsvalues}, both masses turn out to be real and $m_2 > m_0$.

%%%%%%%%%%%%%%%%%%%%%%%%%
\section{Analytic fits}
\label{sec:Fits}

Here we present analytic fits to the numerical data which could be used e.g.\@ for phenomenological applications.
Considering $\gamma_h^{(4)}(p)$, the simple expression
\begin{align}
\gamma^{(4)}_{h,\text{simple}}(p^2) = \frac{G_4^*}{G_4^* + p^2} \,,
\label{eq:G4(p)simple}
\end{align}
with $G_4^* = 0.0360$ captures the correct IR and UV asymptotes and could be used for applications in these limits.
As a more elaborate model we use a Pad\'e - type function
\begin{align}
\gamma^{(4)}_{h, \text{fit}}(p^2) = \frac{a_1+a_2 p^2+a_3 p^4+a_4 p^6+a_5 p^8}{b_1+b_2 p^2+b_3 p^4+b_4 p^6+b_5 p^8+b_6 p^{10}} \,,
\label{eq:G4(p)Fit}
\end{align}
with free parameters $a_i$ and $b_i$. The best-fit values for this model are given by
\begin{align}
a_1&=0.0075542 \,, & \quad a_2&=7.3967 \,, &\quad a_3&=109.42 \,, \nonumber\\[1ex]
a_4&=122.40 \,, & \quad a_5&=8.4140 \,, &  \nonumber\\[1ex]
 b_1&=0.0075403 \,, & \quad b_2&=7.4773 \,, & \quad b_3&=153.00 \,, \nonumber\\[1ex]
b_4&=876.09 \,, & \quad b_5&=2261.9 \,, & \quad b_6&=228.87 \,.
\label{eq:fitparametersG4}
\end{align}
The fit describes the data well in the entire momentum range where numerical data is available with largest deviation from the data of $\sim 2 \%$.

The best-fit values for the simple analytic model of the form factors \labelcref{eq:formfactorSimpleAnalytic} are given by
\begin{align}
p_0 \approx 0.091916 \,, \quad p_1 \approx 0.27551 \,,
\label{eq:fitparametersSimpleAnalytic}
\end{align}
expressed in units of the Planck mass $p_i/M_{\text{Pl}}$. The Planck scale is defined dynamically by ${M_{\text{Pl}}^{2} \coloneqq \gamma_g^{(3)}(0)}$.

As a more elaborate fit function we employ
\begin{align}
f_{R_{\mu\nu}^2\text{, fit}}(p^2) = a_0 + \displaystyle\sum_{i=1}^{4}\frac{a_i}{\left(p/p_i\right)^2+1} \,,
\label{eq:FitFuncfRmn}
\end{align}
for $f_{R_{\mu\nu}^2}(p^2)$ with best-fit values
\begin{align}
a_0&=-0.023601 \,, & \, a_1&=-0.13727 \,, &\, a_2&=0.13138 \,, \nonumber\\[1ex]
a_3&=-0.22100 \,, & \, a_4&=-0.15080 \,, &\, p_1&=0.12436 \,, \nonumber\\[1ex]
p_2&=1.2476 \,, & \, p_3&=0.56405 \,, &\, p_4&= 0.021230 \,.
\label{eq:fitparametersfRmn}
\end{align}
For $f_{R^2}(p^2)$ we use
\begin{align}
f_{R^2, \text{fit}}(p^2) = a_0 + \displaystyle\sum_{i=1}^{3}\frac{a_i}{\left(\left(p/p_i\right)^2+1\right)^2} \,,
\label{eq:FitFuncfR2}
\end{align}
with best-fit values given by
\begin{align}
a_0&=0.028373 \,, & \quad a_1&=0.012637 \,, &\quad a_2&=1.2661 \,, \nonumber\\[1ex]
a_3&=0.57040 \,, & \quad p_1&=5.7131 \,, &\quad p_2&=0.73200 \,, \nonumber\\[1ex]
p_3&=0.092956 \,. & \quad & &\quad &
\label{eq:fitparametersfR2}
\end{align}
The fits describe the data to an accuracy of a few percent, with the largest deviations around $p\approx 1$.

%%%%%%%%%%%%%%%%%%%%%%
\bibliography{references}

\end{document}